\documentclass[twoside,twocolumn,9pt]{article}
\usepackage{extsizes}
\usepackage[super,sort&compress,comma]{natbib} 
\usepackage[version=3]{mhchem}
\usepackage[left=1.5cm, right=1.5cm, top=1.785cm, bottom=2.0cm]{geometry}
\usepackage{balance}
\usepackage{mathptmx}
\usepackage{sectsty}
\usepackage{graphicx} 
\usepackage{lastpage}
\usepackage[format=plain,justification=justified,singlelinecheck=false,font={stretch=1.125,small,sf},labelfont=bf,labelsep=space]{caption}
\usepackage{float}
\usepackage{fancyhdr}
\usepackage{fnpos}
\usepackage[english]{babel}
\addto{\captionsenglish}{}
\usepackage{array}
\usepackage{droidsans}
\usepackage{charter}
\usepackage[T1]{fontenc}
\usepackage[usenames,dvipsnames]{xcolor}
\usepackage{setspace}
\usepackage[compact]{titlesec}
\usepackage{hyperref}
\usepackage{epstopdf}%This line makes .eps figures into .pdf - please comment out if not required.
\definecolor{cream}{RGB}{222,217,201}
\usepackage{dcolumn} 
\usepackage{bm} 
\usepackage{amssymb} 
\usepackage{placeins}
\usepackage{amsmath} 
\usepackage{footmisc}
\usepackage{braket}
\usepackage[latin1]{inputenc}
\usepackage{tikz}
\usetikzlibrary{shapes,arrows}

\newcommand{\crG}[1]{\hat{c}_{#1,\sigma}^{\dagger}}

\newcommand{\deG}[1]{\hat{c}_{#1,\sigma}}
\newcommand{\nuU}[1]{\hat{n}_{#1,\uparrow}}
\newcommand{\nuD}[1]{\hat{n}_{#1,\downarrow}}
\newcommand{\nuG}[1]{\hat{n}_{#1}}
\newcommand{\RA}{\rangle}
\newcommand{\LA}{\langle}

 \newcommand{\lsi}{Laboratoire des Solides Irradi\'es, \'Ecole Polytechnique, CNRS, CEA/DRF/IRAMIS, Institut Polytechnique de Paris, F-91128 Palaiseau, France.}
 \newcommand{\etsf}{European Theoretical Spectroscopy Facility (ETSF).}
\newcommand{\soleil}{Synchrotron SOLEIL, L'Orme des Merisiers, Saint-Aubin, BP 48, F-91192 Gif-sur-Yvette, France.}
\newcommand{\polito}{Politecnico di Torino, 10129 Torino, Italy.}
\newcommand{\UPsud}{Universit\'e Paris-Saclay, 91405, Orsay, France.}
\begin{document}
\pagestyle{fancy}
\thispagestyle{plain}
\fancypagestyle{plain}{
%%%HEADER%%%
\renewcommand{\headrulewidth}{0pt}}
%%%END OF HEADER%%%
%%%PAGE SETUP - Please do not change any commands within this section%%%
\makeFNbottom
\makeatletter
\renewcommand\LARGE{\@setfontsize\LARGE{15pt}{17}}
\renewcommand\Large{\@setfontsize\Large{12pt}{14}}
\renewcommand\large{\@setfontsize\large{10pt}{12}}
\renewcommand\footnotesize{\@setfontsize\footnotesize{7pt}{10}}
\makeatother
\renewcommand{\thefootnote}{\fnsymbol{footnote}}
\renewcommand\footnoterule{\vspace*{1pt}% 
\color{cream}\hrule width 3.5in height 0.4pt \color{black}\vspace*{5pt}} 
\setcounter{secnumdepth}{5}
\makeatletter 
\renewcommand\@biblabel[1]{#1}            
\renewcommand\@makefntext[1]% 
{\noindent\makebox[0pt][r]{\@thefnmark\,}#1}
\makeatother 
\renewcommand{\figurename}{\small{Fig.}~}
\sectionfont{\sffamily\Large}
\subsectionfont{\normalsize}
\subsubsectionfont{\bf}
\setstretch{1.125} %In particular, please do not alter this line.
\setlength{\skip\footins}{0.8cm}
\setlength{\footnotesep}{0.25cm}
\setlength{\jot}{10pt}
\titlespacing*{\section}{0pt}{4pt}{4pt}
\titlespacing*{\subsection}{0pt}{15pt}{1pt}
%%%END OF PAGE SETUP%%%
%%%FOOTER%%%
\fancyfoot{}
%\fancyfoot[LO,RE]{\vspace{-7.1pt}\includegraphics[height=9pt]{head_foot/LF}}
%\fancyfoot[CO]{\vspace{-7.1pt}\hspace{13.2cm}\includegraphics{head_foot/RF}}
%\fancyfoot[CE]{\vspace{-7.2pt}\hspace{-14.2cm}\includegraphics{head_foot/RF}}
%\fancyfoot[RO]{\footnotesize{\sffamily{1--\pageref{LastPage} ~\textbar  \hspace{2pt}\thepage}}}
%\fancyfoot[LE]{\footnotesize{\sffamily{\thepage~\textbar\hspace{3.45cm} 1--\pageref{LastPage}}}}
\fancyhead{}
\renewcommand{\headrulewidth}{0pt} 
\renewcommand{\footrulewidth}{0pt}
\setlength{\arrayrulewidth}{1pt}
\setlength{\columnsep}{6.5mm}
\setlength\bibsep{1pt}
%%%END OF FOOTER%%%
%%%FIGURE SETUP - please do not change any commands within this section%%%
\makeatletter 
\newlength{\figrulesep} 
\setlength{\figrulesep}{0.5\textfloatsep} 
\newcommand{\topfigrule}{\vspace*{-1pt}% 
\noindent{\color{cream}\rule[-\figrulesep]{\columnwidth}{1.5pt}} }
\newcommand{\botfigrule}{\vspace*{-2pt}% 
\noindent{\color{cream}\rule[\figrulesep]{\columnwidth}{1.5pt}} }
\newcommand{\dblfigrule}{\vspace*{-1pt}% 
\noindent{\color{cream}\rule[-\figrulesep]{\textwidth}{1.5pt}} }
\makeatother
%%%END OF FIGURE SETUP%%%
\twocolumn[
\begin{@twocolumnfalse}
%{\includegraphics[height=30pt]{head_foot/journal_name}\hfill\raisebox{0pt}[0pt][0pt]{\includegraphics[height=55pt]{head_foot/RSC_LOGO_CMYK}}\\[1ex]
%\includegraphics[width=18.5cm]{head_foot/header_bar}}\par
\vspace{1em}
\sffamily

\begin{center}
\noindent\LARGE{\textbf{Insights into one-body density matrices using deep learning}}%$^\dag$}} 
\\%Article title goes here instead of the text "This is the title"
%\vspace{0.3cm} & \vspace{0.3cm} \\
\end{center}

\noindent\large{Jack Wetherell\textit{$^{a,b\ddag}$}, %$^{\ast}$\textit{$^{a}$} 
                   Andrea Costamagna\textit{$^{c,d,e,b}$}, %\textit{$^{b\ddag}$}, 
                   Matteo Gatti\textit{$^{a,b,c}$},
                   and Lucia Reining\textit{$^{a,b}$}} \\%Author names go here

\noindent\normalsize{The one-body reduced density matrix (1-RDM) of a many-body system at zero temperature gives direct access to many observables, such as the charge density, kinetic energy and occupation numbers. It would be desirable to express it as a simple functional of the density or of other local observables, but to date satisfactory approximations have not yet been found. Deep learning is the state-of the art approach to perform high dimensional regressions and classification tasks, and is becoming widely used in the condensed matter community to develop increasingly accurate density functionals. Autoencoders are deep learning models that perform efficient dimensionality reduction, allowing the distillation of data to its fundamental features needed to represent it. By training autoencoders on a large data-set of 1-RDMs from exactly solvable real-space model systems, and performing principal component analysis, the machine learns to what extent the data can be compressed and hence how it is constrained. We gain insight into these machine learned constraints and employ them to inform approximations to the 1-RDM as a functional of the charge density. We exploit known physical properties of the 1-RDM in the simplest possible cases to perform feature engineering, where we inform the structure of the models from known mathematical relations, allowing us to integrate existing understanding into the machine learning methods. By comparing various deep learning approaches we gain insight into what physical features of the density matrix are most amenable to machine learning, utilising both known and learned characteristics.
} \\%The abstract goes here instead of the text "The abstract should be..."

\end{@twocolumnfalse} \vspace{0.6cm}]
%%%END OF TITLE, AUTHORS AND ABSTRACT%%%
%%%FONT SETUP - please do not change any commands within this section
\renewcommand*\rmdefault{bch}\normalfont\upshape
\rmfamily
\section*{}
\vspace{-1cm}
%%%FOOTNOTES%%%
\footnotetext{\textit{$^{a}$~\lsi}}
\footnotetext{\textit{$^{b}$~\etsf}}
\footnotetext{\textit{$^{c}$~\soleil}}
\footnotetext{\textit{$^{d}$~\polito}}
\footnotetext{\textit{$^{e}$~\UPsud}}
%Please use \dag to cite the ESI in the main text of the article.
%If you article does not have ESI please remove the the \dag symbol from the title and the footnotetext below.
%\footnotetext{\dag~Electronic Supplementary Information (ESI) available: [details of any supplementary information available should be included here]. See DOI: 00.0000/00000000.}
%additional addresses can be cited as above using the lower-case letters, c, d, e... If all authors are from the same address, no letter is required
\footnotetext{\ddag~ Personal email: {jack.wetherell@polytechnique.edu}; Personal webpage: {https://jw1294.github.io/}}

\section{Background and Objectives}
% What is the Many-Body problem, why is it important and why is it hard?
The development of modern technology is driven by our understanding of the behavior of systems at the quantum mechanical level. Theory and numerical calculations play an important role in the development of this understanding. However, real materials consist of interacting particles, which gives rise to the vastly unfavourable computational and memory scaling required to solve the underlying equations. If we could solve the many-body Schr{\"o}dinger equation for the ground-state wavefunction and store such an object, observables could be calculated as expectations values, but this is not possible for systems of interest. The Hohenberg-Kohn theorems within density functional theory (DFT) tell us that we can instead describe any observable in terms of the much more manageable electron density\cite{Hohenberg-Kohn}, but the form of almost all such functionals is unknown.

% What is the density matrix and why is it useful?
The one-body reduced density matrix (1-RDM) can be thought of as an intermediate quantity between these two extremes. As with the density, it avoids the problem of having to store a function of all the spin and spacial coordinates of the system. For a $N$-electron spin-resolved system at zero temperature the 1-RDM is given by 
\begin{equation}
\gamma(r,r') = N \int \Psi(r, r_2, r_3, \dots) \Psi^*(r', r_2, r_3, \dots) dr_2 dr_3 \dots.
\end{equation}
Its diagonal is the charge density $n(r)=\gamma(r,r)$. The expectation value of any local or non-local one-body operator in terms of the density matrix is
\begin{equation}
O[\gamma] = \int  \,O(r,r')\gamma(r,r') dr dr'.
\end{equation}
In particular, the kinetic energy $K$ of the many-body system reads
\begin{equation}
K[\gamma] = -\frac{\hslash^2}{2m}\int \nabla^2 \left. \gamma(r,r') \right| _{r=r'} dr'.
\end{equation}
While reduced density matrix functional theory (RDMFT)\cite{Coleman1963,Gilbert1975,Levy1979,Valone1980,Pernal2016,Lathiotakis2007,Piris2017,Schilling2018,Giesbertz2018,doi:10.1063/1.1906203,PhysRevB.78.201103} performs a constrained minimisation of the total energy $E$ over the 1-RDM, it would be possible to perform the minimisation over the density itself if we could express the 1-RDM as functional of the density. This would allow for direct minimisation of the energy within DFT without the need for a Kohn-Sham (KS) auxiliary system\cite{Kohn-Sham}, which introduces orbitals\cite{PhysRevLett.81.866}. \emph{Therefore it would be highly desirable to find the functional $\gamma[n]$, as this would allow these key quantities to be themselves expressed as functionals of the charge density.} The search for such a functional does not have to be completely blind. In particular, the density matrix is an object that is subject to many constraints\cite{Coleman1963}. Not all functions $f(r,r')$ are valid density matrices, in the sense that they can be computed from the ground state wavefunction of a Hamiltonian with a local and static potential. The knowledge of constraints is crucial when building functionals, as it considerably reduces the domain of legitimate functionals one must search over\cite{doi:10.1063/1.5025668}. 

% What is machine learning and why is it useful?
In the data science community, there is an exponential growth of modern machine learning methods, that each day are being applied to successfully solve increasingly difficult problems with astonishing accuracy. Such problems were previously thought to be impossible to solve numerically, in particular in the field of image processing. As the 1-RDM stored on a numerical grid is essentially an image, with a dominant spacial structure, the question naturally arises: \emph{Can these methods be used to gain new insights into the 1-RDM and help us find the functional we desire?}

% How is machine learning currently used in many-body physics?
Machine learning is becoming increasingly utilised in the field of condensed matter physics\cite{RevModPhys.91.045002,PhysRevLett.108.253002,PhysRevB.94.245129,doi:10.1002/qua.25040,PhysRevLett.98.146401,PhysRevLett.104.136403,moreno2019deep,doi:10.1063/1.4707167,McGibbon2013,McDonagh2018,PhysRevLett.108.058301,Hautier2010,Kolb2017,PhysRevA.100.022512,Schutt2019,Schmidt2019,PhysRevA.96.042113,suzuki2020machine,Nagai2020}. In particular, machine learning has been shown to yield impressive results for the computation of the exchange-correlation potential within DFT. In a recent work\cite{Zhou2019}, small exactly solvable molecules are used to train a machine learning model for the exchange-correlation potential. The authors demonstrate that this can then be used to predict the properties of more complex molecules. This exploits the holographic density principal of molecules\cite{doi:10.1063/1.5012279}, which suggests that the behaviour at a given part of a large molecule (for example a bond) is also present in a small molecule. Machine learning is also widely utilised within condensed matter physics, and has been shown to be able to perform the Hohenberg-Kohn mapping from the external potential to the charge density directly using kernel ridge regression\cite{Brockherde2017}.

% What are our objectives in this work?
We wish to augment machine learning models with our current approaches, such that only the smallest possible parts, which are the most difficult to approximate, have to be learned. This raises three fundamental questions: \emph{Can machine learning give insights to the 1-RDM, in particular constraints? Can machine learning algorithms optimised for image processing learn the functional $\gamma[n]$, and can we integrate this with pre-exiting physically-based models so we need only learn the neglected phenomena, and if so which part is the most amenable to machine learning?}

\section{Machine Learning Methods}
% What is deep learning and how is it important?
Deep learning is a powerful method within machine learning that is used to perform very high dimensional and extremely non-linear fitting using a large data-set on powerful hardware. We now introduce the deep learning methods that we utilise to answer our proposed questions, and how in particular they relate to physical problems faced in quantum chemistry.

\subsection{Deep Neural Networks}
% What problems do deep neural networks solve? 
Deep neural networks are numerical models that are trained to recognise patterns and relationships between data. For our purposes we will use them to perform generalised regression. If we have a labeled data-set of known inputs $\{x \}$ and known outputs $\{y \}$ a deep neural network can learn any non-linear map $f:x \rightarrow y$, that can make predictions on novel $x$ values. This is learned through the process of gradient descent, where the parameters of the network are adjusted to minimise the error of predictions made on known data. With proper structuring the inputs and outputs can be of any form: images, functions, numerical values etc, and the model with enough complexity can learn any arbitrarily non-linear mapping.

\begin{figure}[htbp]
\centering
\includegraphics[width=1.0\linewidth]{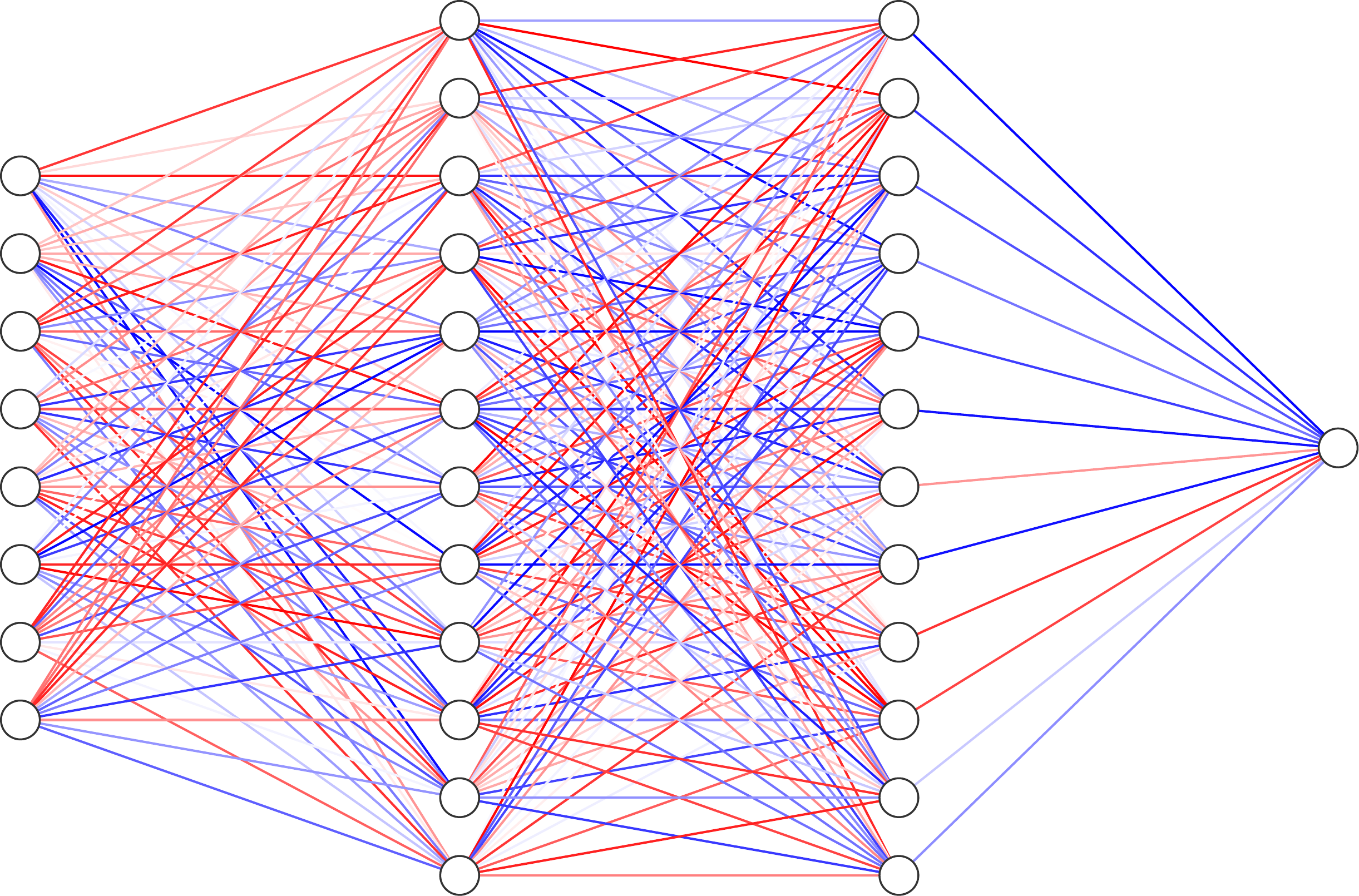}
\caption{An illustration of a multilayer perceptron (MLP)\cite{NNSVG}. The circles represent the layers of perceptrons, that are fully connected between layers. The red and blue lines represent the values of the weights of each layer (one set of  $\{w_i\}$ for each perceptron), where blue indicates a positive weight, and red a negative weight. Each perceptron also has a bias $b$ that is not shown. It is these weights and biases that are adjusted during the training via gradient descent. The middle two layers are termed \emph{hidden layers} as they are not directly connected to the inputs or outputs. With a sufficient number of perceptrons in the hidden layers, this network can in principal learn any arbitrarily complex mapping from the 8 input values, to the 1 output value $y_1 = f(x_1,x_2,\dots, x_8)$.}
\label{fig:mlp}
\end{figure}

% What is a MLP?
The simplest type of neural network we will utilise is the multilayer perceptron (MLP)\cite{mcculloch1943logical}, illustrated in figure \ref{fig:mlp}. A MLP is composed of layers of perceptions, each holding one value, computed as a weighted linear sum of its inputs $\{ x\}$ passed through some non-linear activation function $\sigma$: $\sigma \left( \sum_i w_i x_i + b\right)$, where $\{ w\}$ are the weights of the layer and $b$ is the perception's bias. It is the many layers of these perceptrons, each a non-linear combination of all the perceptrons in the previous layer, that allows the network to learn highly intricate relationships. During training the weights and biases are adjusted through gradient descent, after being randomly initialised, until the error with respect to the known data is minimised. 

\subsection{Autoencoders}
%  What is an Autoencoder?
Autoencoders (AE) are deep neural networks that are trained to perform efficient generalised data compression\cite{liou2014autoencoder}. They consist of a neural network that is trained to reproduce exactly its own input data as output data $f_\mathrm{AE}:x \rightarrow x$. A single hidden layer acts as a bottleneck, containing fewer perceptions than in the input and output layers, compressing the data to a latent space. The two parts of the autoencoder can be separated into the encoder $f_\mathrm{E}:x \rightarrow x'$ and decoder $f_\mathrm{D}:x' \rightarrow x$, where $x'$ has a smaller dimensionality then $x$. 

% Why are they useful
The amount to which data can be compressed depends on its features. For example, images of only faces can be  compressed significantly more than general images, as they are more heavily constrained. Therefore the use of AEs can be thought of as \emph{domain-specific} data compression, as the network learns the underlying features of a specific data-set (domain), and so learns to exploit these in order to achieve a greater degree of compression. In general, there is a deep connection between compression and constraints: the more data is constrained, the more it can be compressed losslessly. We propose training an autoencoder on the 1-RDM in order to inform: to what extent it can be compressed, the nature of the compression, and how one can extract these constraints. In particular, we would hope the AE would learn the 1-RDM $\gamma(r,r')$ can be compressed to a latent space of the dimensionality of its diagonal $n(r)$.

\begin{figure}[htbp]
\centering
\includegraphics[width=1.0\linewidth]{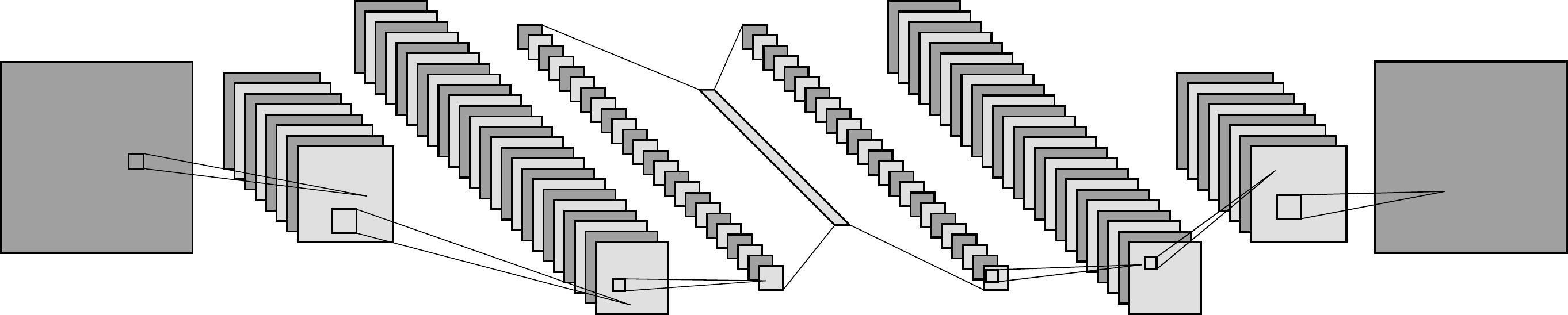}
\caption{Illustration of a convolutional autoencoder (CAE)\cite{NNSVG}. The 2 dimensional input image is convolved with kernels reducing the spacial dimensions and increasing the number of features until the data is totally reduced to a feature latent space with far fewer degrees of freedom that the original image. This compression can be achieved as the image contains some intrinsic structure, as opposed to totally random pixel values. This process is then inverted with transpose convolutional layers until the original image dimension is recovered. This network is trained by adjusting the kernels through gradient descent until the output image reproduces the input image to a given tolerance over the training data-set.}
\label{fig:cae}
\end{figure}

% What is a CAE? 
In principle we could use a MLP with a bottleneck layer as our deep autoencoder. However, this would be onerously expensive and inefficient, for the same reasons MLPs are rarely used for image processing: they do not exploit the spacial structure, treating each pixel of data totally independently from the rest. Instead, density matrices do have strong spacial structure: for example, for the most common external potentials they are continuous and smooth. Therefore, we utilise convolutional autoencoders (CAE) to learn the constraints of the 1-RDM. CAEs convolve several kernels over the two dimensional input image using element wise multiplication\cite{TRIVEDI2018525,Zhang:90}. These values are then passed to some non-linear activation function $\sigma$. This can be thought of as 'scanning' over the image with a filter representing a particular feature. The resultant values describe the similarity between a region of the image and the feature of interest. It is this that exploits the spacial structure of the image. This process is repeated until the spacial information of the image has been converted from real space to a 1-dimensional feature space. This is our bottleneck layer. This process is then reversed using transpose convolutions layers (that perform the inverse operation) until the image is recovered. This is illustrated in figure \ref{fig:cae}. \emph{It is the kernels of this network that are adjusted throughout training}, until the input image can be reconstructed as the output to a required tolerance over the data-set. This then allows us to learn arbitrarily non-linear constraints of the 1-RDM, and find a latent feature space where the 1-RDM as a functional of the density may be simpler. We employ CAEs to learn constraints and develop approximate functionals for the 1-RDM on a large data set.

\subsection{Principal Component Analysis}
\label{sec:pca}
\begin{figure}[htbp]
\centering
\includegraphics[width=1.0\linewidth]{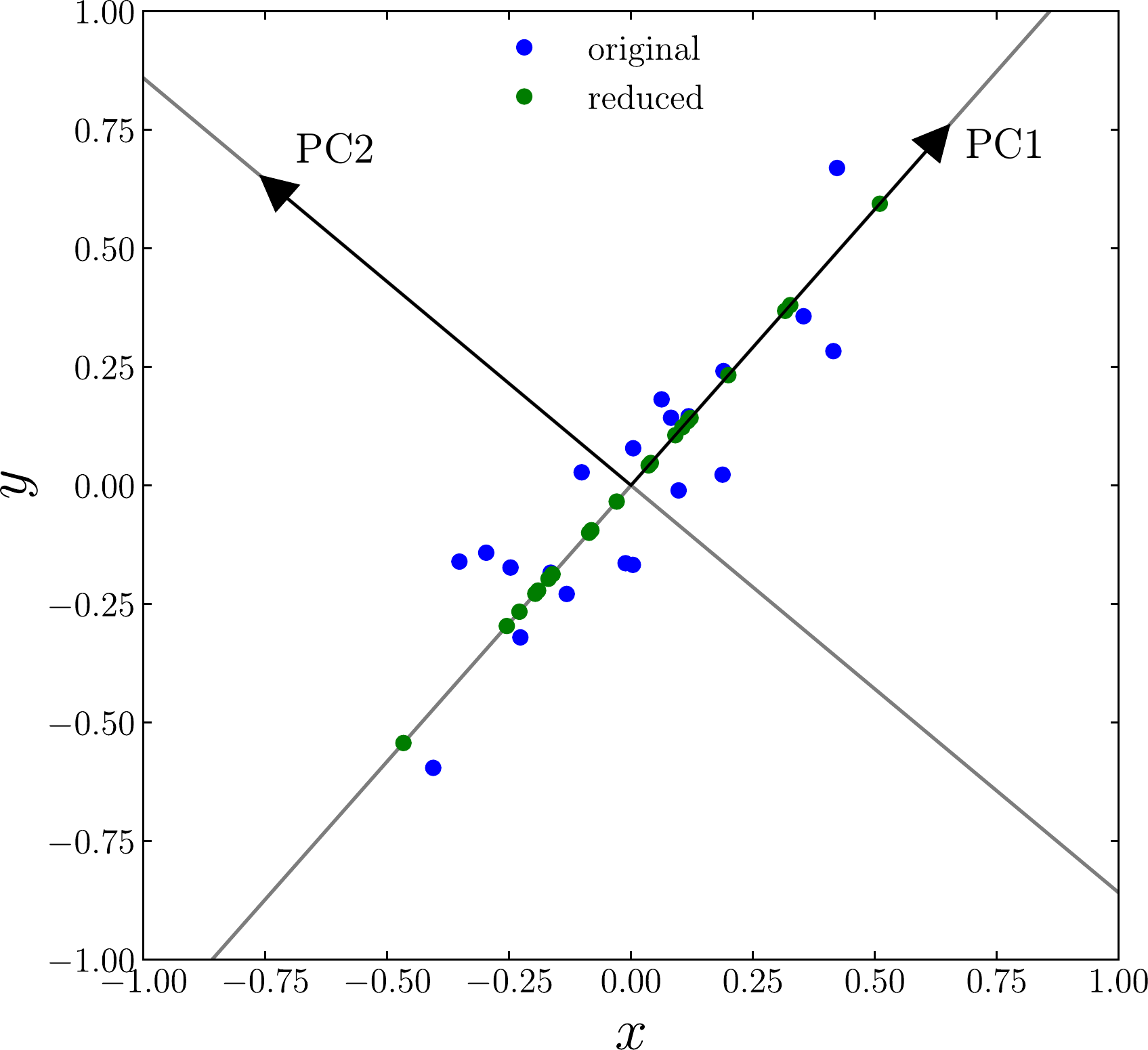}
\caption{A simple illustration of principal component analysis (PCA). The blue dots show a set of paired data points $\{(x_i, y_i)\}$. The PCA applied to this data yields an orthonormal basis shown by the two black arrows. They are ordered by their variance, with principal component 1 (PC1) being the component of most variation. If we then use this to perform a lossy compression we simply discard the principal component 2 (PC2) and perform the inverse transform, yielding the reduced data points shown in green.}
\label{fig:illustrate_pca}
\end{figure}

%  What is PCA?
The simplest autoencoder we can imagine is dimensional reduction via principal component analysis (PCA)\cite{Jolliffe2011}. PCA consists of computing the linear transform to an orthogonal space that is designed such that each component is ordered by its variance \cite{pearson1901liii}. This is illustrated in figure \ref{fig:illustrate_pca}. If the variance of a given component is zero, that component can be neglected such that the original data is recovered exactly upon the inverse transformation. Good approximations are obtained when components are neglected whose variance is small. It is important to note that the PCA is a strictly linear transformation, and so can only determine linear constraints in data (in contrast to CAEs).

% Mathematical description of PCA
We will now introduce how PCA is performed on a data-set consisting of  $T$ $N$x$N$ matrices. We begin by considering element $t$ of our data-set:
\begin{equation}
{\gamma^{(t)}}=\begin{bmatrix}
\gamma_{N1}^{(t)}& \cdots & \gamma_{NN}^{(t)}\\
\vdots & \ddots & \vdots \\
\gamma_{11}^{(t)} & \dots & \gamma_{1N}^{(t)}
\end{bmatrix}\in \mathbb{R}^{N\times N}.
\end{equation}
In order to represent this matrix, it is possible to define a $N^2$-dimensional basis, each component of which points to a different entry of the matrix $$\mathcal{B}_e=\{|e_r\RA\}_{r=1}^{N^2}=\{|e_{r[i,j]}\RA\}_{i,j=1}^{N}$$
This basis leads to the `flattened' version of the original matrix, represented as the following vector:
\begin{equation}
|\gamma^{(t)}\RA = \sum_{i,j=1}^{N}\LA e_{r[ij]} |\gamma^{(t)}\RA\cdot|e_{r[ij]}\RA = \sum_{i,j=1}^{N} \gamma_{ij}^{(t)} |e_{r[ij]}\RA
\end{equation}
or, also
\begin{equation}
\underline{\gamma}^{(t)}=
\begin{bmatrix}
 \gamma_{11}^{(t)}& \gamma_{12}^{(t)} & \cdots & \gamma_{NN}^{(t)} 
\end{bmatrix}\in \mathbb{R}^{N^2}.
\end{equation}
Out data-set of $T$ such vectors is then denoted:
\begin{equation}
{\Gamma}=
\begin{bmatrix}
\underline{\gamma}^{(1)}\\
\vdots\\
\underline{\gamma}^{(T)}
\end{bmatrix}
\end{equation}
The disposal of this data-set allows us to define a new basis with which it is possible to describe the $\gamma$-vectors. The PCA is then considered a linear numerical method to determine the following two sets of quantities:
\begin{itemize}
\item $|\gamma_0\RA$: The mean matrix. The knowledge of this allows writing each matrix under analysis in terms of its variations from the mean $|\gamma\RA = |\gamma_0\RA+|\Tilde{\gamma}\RA$. This is termed the mean-adjusted matrix. The mean matrix components in the previously defined basis read
$$(\gamma_0)_{ij}=\frac{1}{T}\sum_{t=1}^T\gamma_{ij}^{(t)}.$$
\item $\mathcal{B}_{pca}=\{|p_i\RA\}_{i=1}^{N^2}$: A new basis, corresponding to the principal components (or principal directions). They are the normalized eigenvectors of the covariance matrix 
$$C =\frac{1}{T-1}{\Gamma}^{\dagger}{\Gamma} = \sum_{r,r'=1}^{N^2}c_{r,r'}|e_r\RA\LA e_{r'}|$$ $$c_{r,r'} = \frac{1}{T-1}\sum_{t=1}^T(\gamma_r^{(t)}-(\gamma_0)_r)(\gamma_{r'}^{(t)}-(\gamma_0)_{r'}).$$ The eigenvalues of such a matrix are termed the variances $\{\sigma_i^2\}_{i=1}^{N^2}$. The principal components are the directions along which, in the data-points, there are the most informative variations with respect to the average matrix (see figure \ref{fig:illustrate_pca}). They are sorted by importance depending on the value of the associated eigenvalue. 
\end{itemize}
The knowledge of the data-set in this form implies that, by solving the eigenequation, one can determine the coefficients $\LA e_r|p_i\RA$ in the expansion
\begin{equation}
|p_i\RA = \sum_{r=1}^{N^2}\LA e_r|p_i\RA|e_r\RA.
\end{equation}
Each 1-RDM can be expressed in this new basis in an expansion called \emph{principal components decomposition}:
\begin{equation}
|\gamma\RA =|\gamma_0\RA+ \sum_{i=1}^{N^2}\LA p_i |\Tilde{\gamma}\RA\cdot|p_i\RA.
\end{equation}
The main property of PCA is that the existence of linear constraints in between the features of the object under analysis (entries of the matrix) leads to vanishing eigenvalues, associated to non-informative principal components. 
This allows the compression of the information by using a number $\nu < N^2$ of components. For example, if the matrix is symmetric $\gamma^{(t)}_{i,j} = \gamma^{(t)}_{j,i} \forall t$, the data can be compressed to $\nu \leq \frac{N(N+1)}{2}$, and the matrix can be fully represented using a reduced number of principal components
\begin{equation}\label{eq:pca_trunc}
|\gamma\RA =|\gamma_0\RA+ \sum_{i=1}^{\nu}\LA p_i |\Tilde{\gamma}\RA\cdot|p_i\RA.
\end{equation}

\section{The data-set}
\label{sec:dataset}
% Why do we want a data-set and how we get it
In order to investigate to what extent deep learning can answer our questions of interest, we construct a large training and testing data-set of external potentials, charge densities and 1-RDMs. To generate the data-set we use the \texttt{iDEA} code\cite{PhysRevB.88.241102,PhysRevB.99.045129}. This exactly solves the many-body Schr{\"o}dinger equation for finite systems of up to four electrons interacting via a softened Coulomb interaction on a one-dimensional real-space grid given any arbitrary local external potential. In addition, it provides implementations of many widely-used approximate methods\cite{PhysRevMaterials.2.040801}. After computing the exact many-body wavefunction, any required observables can be obtained via expectation values directly. The model systems solved by the \texttt{iDEA} code have in the past been used to develop improved approximations to DFT\cite{PhysRevB.90.241107, PhysRevA.101.032502}, many-body perturbation theory\cite{PhysRevB.97.121102}, as well as investigating the nature of exact potentials\cite{PhysRevB.93.155146}, where the model systems have been shown to well describe crucial features as that of real three-dimensional molecules\cite{hodgson2017interatomic}.

% Specification of the data-set
The training data is composed of a large family of randomly generated two-electron systems in their spin-resolved ground-state. For each system we: construct a randomly generated smooth potential $V(x)$ for which we determine the exact ground-state many-body wavefunction. From this we compute the charge density $n(x)$ and 1-RDM $\gamma(x,x')$. We also, for the same potential, compute the charge density and 1-RDM using purely non-interacting electrons (NON) and unrestricted Hartree-Fock (UHF). As these are finite systems in the ground state, the 1-RDMs are real-valued functions. We define the external potential as a sum of randomly distributed Fourier components within a large confining potential\cite{Skelt2018}: 
\begin{equation}
\label{eq:sys}
V(x) = D x^{10} + T \sum^{N}_{n=1}\left( a_n \mathrm{cos}\left( \frac{n \pi x}{L} \right)  + b_n \mathrm{sin}\left( \frac{n \pi x}{L} \right)\right),
\end{equation}
where $L=15 \mathrm{(a.u.)}$\footnote{Hartree atomic units: $m_e = \hslash = e = 4\pi\varepsilon_0 = 1$.} is the width of system, $N=3$ is the number of Fourier terms, $D=10^{-11}$ is the damping factor of confining term and $T=0.1$ is damping factor of Fourier terms. $a_n$ and $b_n$ are generated randomly with a uniform distribution from $-\frac{2L}{3}$ to $\frac{2L}{3}$. We generated a data-set of 50,000 systems for 2, 3, 4, 6, and 62 grid points. 

\begin{figure*}[htbp]
\centering
\includegraphics[width=1.0\linewidth]{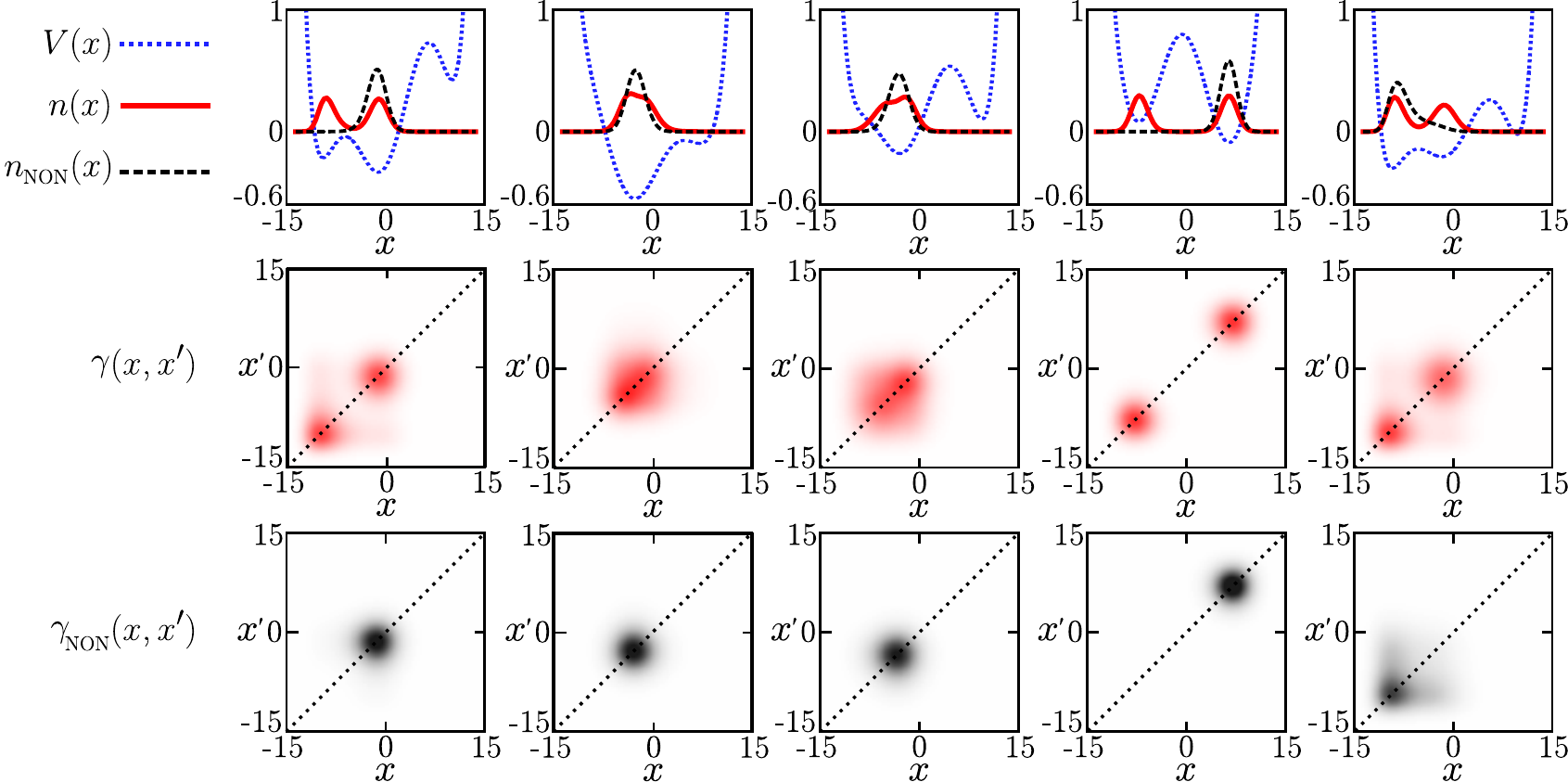}
\caption{The first 5 elements of the 50,000 systems in the data-set. Row 1 shows the randomly generated (as defined by equation \ref{eq:sys}) external potential $V(x)$, the interacting charge density $n(x)$, and purely non-interacting charge density $n_\mathrm{NON}(x)$. These potentials give rise to a wide range of density shapes, locations and overlaps. Row 2 shows each of the systems exact 1-RDM $\gamma(x,x')$, where the dotted lines indicate the diagonal $x=x'$. This can be contrasted with row 3, showing the purely non-interacting 1-RDM $\gamma_\mathrm{NON}(x,x')$. All quantities are given in a.u.}
\label{fig:data}
\end{figure*}

% Exploring the data
Figure \ref{fig:data} shows the first five elements of the 50,000 test systems in the data-set with 62 grid points. The systems display a wide range of potentials, densities and 1-RDMs, exhibiting a wide range of localisation and correlation.
We use this data-set to train and test our deep learning models.

\section{Learning Constraints}
\subsection{Constraints of the charge density} \label{sec:PCAcharge}
\begin{figure}[htbp]
\centering
\includegraphics[width=1.0\linewidth]{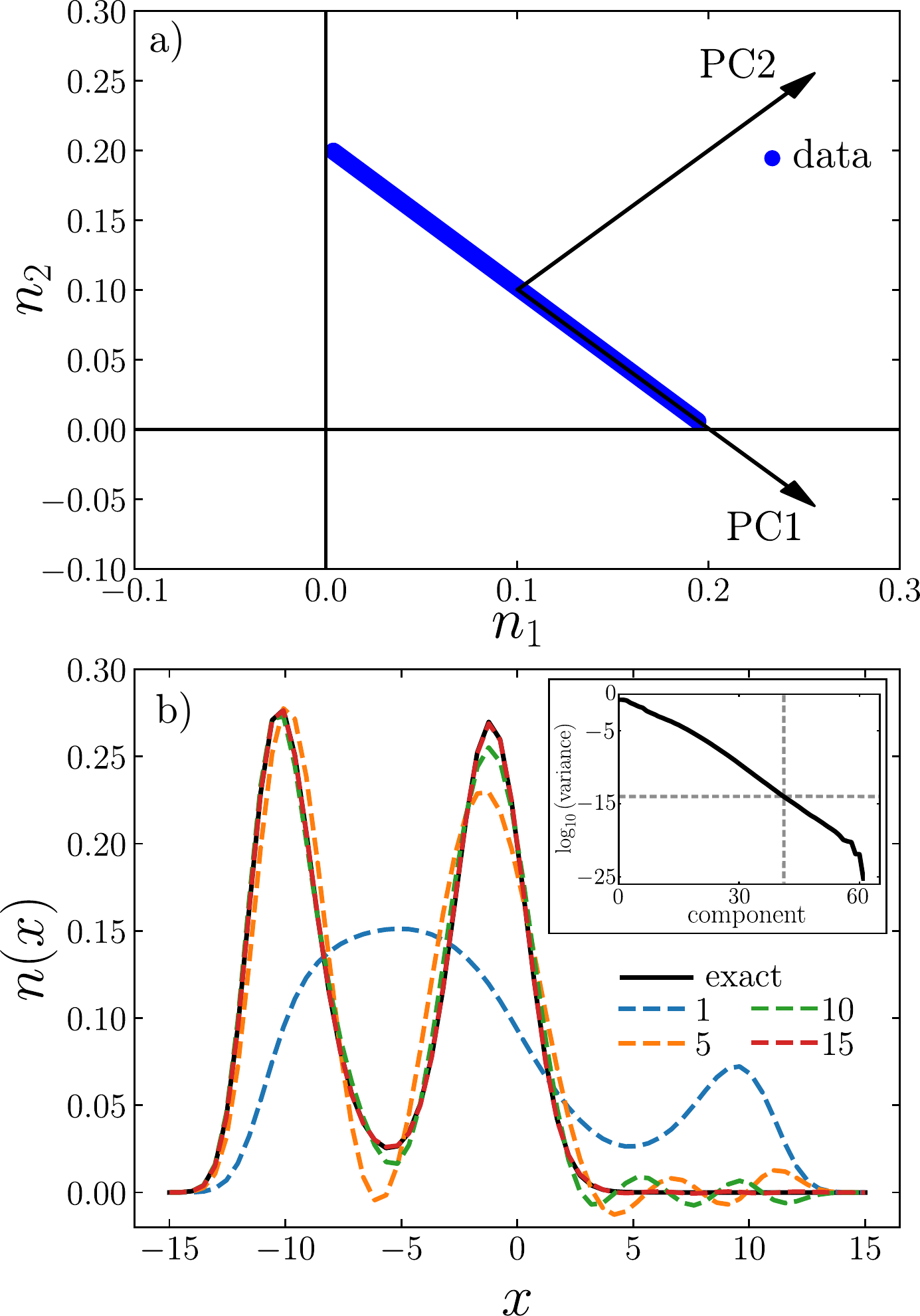}
\caption{PCA being applied to the charge density. Panel (a) shows the density data values of the 2-point data-set along with the two orthogonal principal components (PC). PC1 corresponds to the charge moving between the two points, and PC2 corresponds to changing the net value of charge. From the variances $[2.19 \times 10^{-3}, 4.19 \times 10^{-16}]$ we see that the PCA has learned that the density is constrained by the total charge. By looking at the data this way we can see clearly that this is a linear constraint that the PCA can capture exactly. The inset in panel (b) shows the logarithm (to the base 10) of variance of the PCA components for the 62-point data-set. The horizontal grey dotted line illustrates floating point numerical precision, and the vertical indicates the 42 components needed to obtain such accuracy. Panel (b) shows an example 62-point density of various numbers of included principal components, along with the exact density for comparison.}
\label{fig:n_pca}
\end{figure}

% Applying a PCA to the 2 point density, learns normalisation and positivity.
We will now investigate to what extent the machine can learn fundamental principles. Not all functions of two variables $f(x,x')$ are 1-RDMs due to its constraints, and so we now investigate to what extent the machine can learn such constraints. To begin we will focus on the simplest example possible. We will use the data-set of only 2-points, as it makes it possible to visualise and quantify all the relationships between the data. We will first see if we can use PCA to learn the known constrains of the \emph{charge density}. As the space only contains 2 points, the density is represented by 2 values $n_1 = n(x_1)$, $n_2 = n(x_2)$. The following are two known constraints:
\begin{enumerate}
\item $n_1 + n_2 = \frac{N}{\Delta x}$
\item $n_1 > 0$ and $n_2 > 0$.
\end{enumerate}
Applying PCA to the data-set we find the components shown in figure \ref{fig:n_pca}(a). The principal components have the variances  $[2.19 \times 10^{-3}, 4.19 \times 10^{-16}]$. The PCA has encoded some physical insight: the first principal component describes that if an amount of charge is removed from one spacial position, it must be added to the other spacial position. This describes the charge is free to move along the $x$-axis. Component 2 describes adding and removing charge from the system. As the variance of this component is (numerically) zero, it indicates that the amount of charge must be the same as the average system, therefore illustrating the conservation of particle number. The range of these components in the data-set gives us the positivity condition. This shows that in this simple case, PCA can be used to encode both of the constraints on the density, due to their linearity. We see this trend continues to the 3, 4, 6, and 62 grid point data-sets: the PCA finds that a $N$ point system can be reduced to \emph{at least}
\begin{equation}
N \rightarrow N-1
\end{equation}
components losslessly, as the final component is entirely determined by the linear normalisation condition. Therefore the points of the density lie in a $N$ dimensional-flat plane.

% Applying a PCA to the 62 point density, learns additional constraints from structure.
When considering the solution to a quantum system we assume $V(x)$ can take any form that gives a valid solution of the Schr\"odinger equation. But in reality, when studying a class of systems, such as molecules, the range of external potentials is much smaller, simply determined from the atomic positions and charges. These potentials are constrained by the atomic nature of matter within the Born-Oppenheimer framework. Constraining $V(x)$ in this way has the effect of also constraining the observables - maybe some of these constraints are linear. If there are additional linear constraints, they will be found by PCA. Our 62-point data-set has a characteristic well defined structure, as it is formed from a Fourier series and confining potential, in addition to the usual smoothness and continuous requirements of the charge density. This is in contrast to the 2 point case where potential is essentially 2 independent random values, where there is no concept of smoothness. The inset of figure \ref{fig:n_pca} (b) shows the logarithm (to the base 10) of the variance of each component in the 62-point case. This shows that only 41 components are necessary to describe the density to numerical accuracy, much smaller than $N-1$ (61). This has captured these additional linear constraints. Figure \ref{fig:n_pca} (b) shows how the structure of the density is assembled from adding successive principal components. Only 15 are needed to reduce the error to $10^{-6}$ (a.u.). This illustrates that using a constrained class of external potentials, in addition to the usual smoothness constraints, leads to additional constraints in the charge density, which in turn leads to additional linear constraints that can be extracted using PCA.

\subsection{Constraints of the density matrix} \label{sec:pcadm}
\begin{figure}[htbp]
\centering
\includegraphics[width=1.0\linewidth]{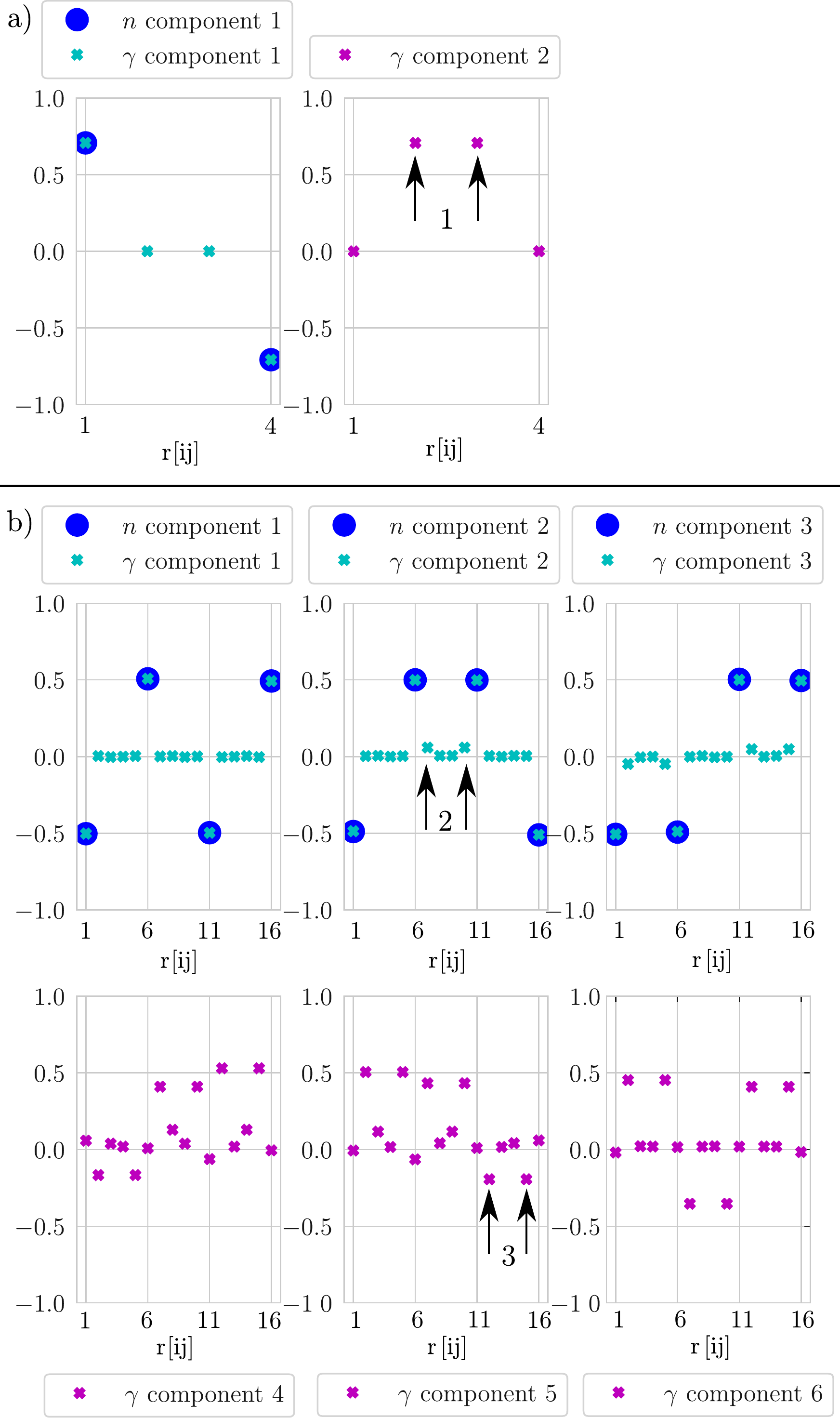}
\caption{PCA being applied to the 'flattened' (see equation \ref{eq:flat}) 1-RDM. Panel (a) shows a comparison of the non-zero variance 1-RDM components with that of the density for the 2-point data-set. \emph{The vertical grey lines indicate the diagonal elements}. For the first principal component the 1-RDM is identical to that of the density along the diagonal and zero value off-diagonal. The second principal component shows that the upper and lower off-diagonal elements must always equal (symmetry constraint indicated with arrow pair 1). Panel (b) shows a comparison of the first 6 non-zero variance 1-RDM principal components in comparison to the 3 non-zero variance density principal components for the 4-point data-set. \emph{Again, the vertical grey lines indicate the four diagonal elements.} The principal components of the density again match that of the diagonal of the first 3 principal components of the 1-RDM. While components 1 is zero for the off diagonal elements, components 2 and 3 have some small contribution to the off diagonal elements, for example the values indicated by arrow pair 2 (see discussion in main text). The next three components have non-zero value in the off diagonal directions, and the figure illustrates that $\gamma_{ij}=\gamma_{ji}$ (see for example arrow pair 3).}
\label{fig:p_pca1}
\end{figure}

% Applying a PCA to the 2 point density matrix, learns symmetry.
We now apply the PCA to the 1-RDM. We would expect the PCA to learn the same linear constraints as for the density, as the 1-RDM contains the density along its diagonal. Moreover, the PCA should learn the additional linear constraint of symmetry, so altogether:
\begin{enumerate}
\item $\sum_{i} \gamma_{ii} = \frac{N}{dx}$
\item $\gamma_{ii} > 0$ $\forall i$
\item $\gamma_{ij} = \gamma_{ji}$ $\forall i,j$.
\end{enumerate}
Where again we consider the  2-point case, and so the 1-RDM is represented by 4 values $\gamma_{11} = \gamma(x_1, x_1)$, $\gamma_{12} = \gamma(x_1, x_2)$, $\gamma_{21} = \gamma_{21}(x_2, x_1)$, $\gamma_{22} = \gamma(x_2, x_2)$. We will write these in `flattened' form.
\begin{equation}
\label{eq:flat}
\gamma_{ij} \rightarrow \gamma_{r[ij]}.
\end{equation}
In this way, the 1-RDM becomes a $4$-dimensional vector, and so, in our case, the four elements of the 1-RDM are denoted $\gamma_{11} \rightarrow \gamma_{r=1}$, $\gamma_{12} \rightarrow \gamma_{r=2}$, $\gamma_{21} \rightarrow \gamma_{r=3}$, $\gamma_{22} \rightarrow \gamma_{r=4}$. Where  $\gamma_{r=1}$ and  $\gamma_{r=4}$ are the diagonal elements. We apply PCA to this 2-point data-set of 1-RDMs. We observe that this yields 2 components with non-zero variance. These are shown in figure \ref{fig:p_pca1} (a), and compared to the components of the density obtained in section \ref{sec:PCAcharge}. We see that the component 1, corresponding the direction of maximum variance in the data-set, is exactly the same as the corresponding density component, with no non-zero value in the off-diagonal terms. This can be thought as moving along the flat density plane. It is this term (along with the fact that the component changing the net charge has a variance of zero) that captures the first 2 constraints, as in section \ref{sec:PCAcharge}. Component 2 has no non-zero values in the diagonal, but only values in the off-diagonal terms. As indicated by arrow pair 1, the two off-diagonal terms have the same value, this has captured the symmetric constraint. The PCA describes: if you set $\gamma_{12}$ by a given value, you must set $\gamma_{21}$ to exactly the same value.

% Applying a PCA to the 4 point density matrix, learns symmetry, cannot learn non-linearity. 
We now apply the PCA to the 4-point data-set of 1-RDM. This will inform what compression the PCA can perform losslessly: \emph{can it encode the $N^2$ elements of the 1-RDM by only $N$, as in DFT?} We find that this is not the case, as in the 4-point case the PCA can perform the lossless compression from $4^2$ elements to $9$. In general we find that for an $N$-point system the PCA can perform lossless compression to \emph{at least} 
\begin{equation}
N^2 \rightarrow \frac{N(N+1)}{2} - 1. 
\end{equation}
This is simply the number of diagonal elements subtract 1 plus the half the number of off-diagonal elements. This is exactly the amount that is derived from the three constraints of the 1-RDM, and hence the PCA finds there are no additional linear constraints we were missing. Figure \ref{fig:p_pca1} (b) shows the first six non-zero principal components of the 1-RDM. The diagonal values of the first three components correspond exactly to that of the density principal components, and the off-diagonal values are almost zero, except for small features appearing in the elements adjacent to diagonal ones, for example as indicated by arrow pair 2. The remaining components describe only the off-diagonal elements, and once again, due to values coming in pairs (see for example, arrow pair 3), reflect the symmetry constraint. The fact that some off diagonal values are non-zero in the components that correspond to the density is significant as it allows the separation of the linear and non-linear terms of $\gamma[n]$ in a domain specific way. This idea will be developed further in section \ref{sec:PCAfunc}.

\begin{figure}[htbp]
\centering
\includegraphics[width=1.0\linewidth]{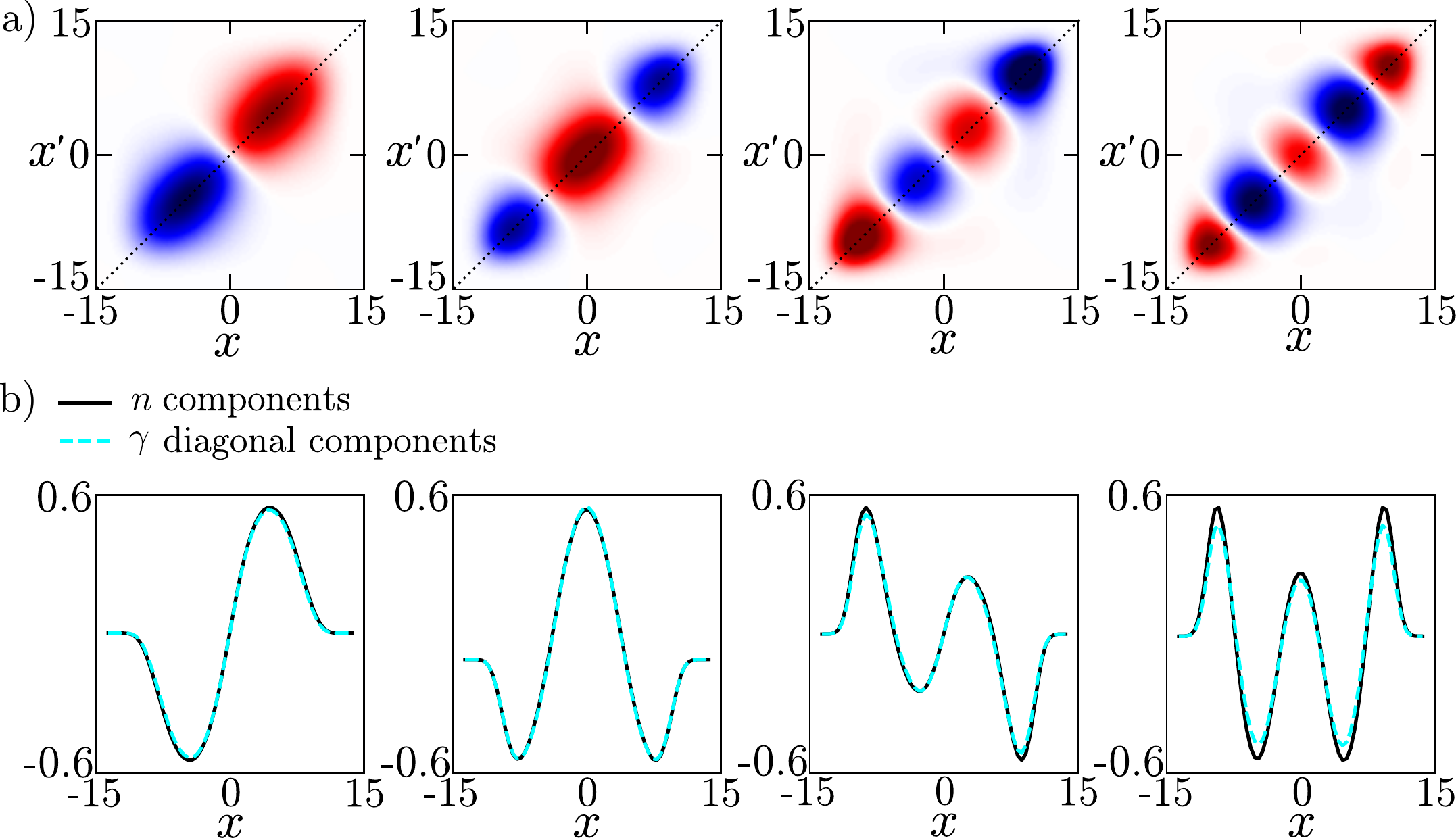}
\caption{PCA being applied to the 1-RDM in the 62-point case. Panel (a) is the first 5 principal components of the 1-RDM for the 62-point data-set. Panel (b) compares the (scaled due to the PCA normalisation convention) diagonals of these components to the first 5 principal components of the density.}
\label{fig:p_pca2}
\end{figure}

% PCA for the 62 point case. New linear constraints.
As we found in section \ref{sec:PCAcharge} that additional linear constraints on the density emerge when the structural constraints are applied to the external potential, and due to the smoothness of the density, we would like to see to what extent this extends to the 1-RDM, and to what extent this can be utilised. The top row of Figure \ref{fig:p_pca2} (c) shows a 2D-view of the first 5 components of the 1-RDM for the 62-point data-set. We would expect that, if there were no additional linear constraints, PCA would find $62^2 \rightarrow \frac{62(62+1)}{2} - 1 = 1952$ lossless compression to be obtained. We find only 327 are required within our numerical precision. This implies that, as we approach the continuum by increasing the number of grid points, additional linear constraints manifest in the 1-RDM. This is because each of the elements $\gamma_{ij}$ cannot be treated independently, there is an emerging additional structure due to the smoothness and continuous properties of the wavefunction, and from the constraints we impose on the external potential being formed from Fourier components. These properties have no meaning in the 2 point system, and hence do not appear. In the bottom row of \ref{fig:p_pca2} (c) we compare the diagonals of these first 5 components (scaled due to the PCA normalisation convention), with the first 5 density components: we see they correspond exactly, but have significant weights off the diagonal elements $\gamma_{ii}$. This allows us to describe the linear part of the functional $\gamma[n]$ using our data-set. We will explore constructing functionals from this premise in section \ref{sec:PCAfunc}. 

\begin{figure*}[htbp]
\centering
\includegraphics[width=1.0\linewidth]{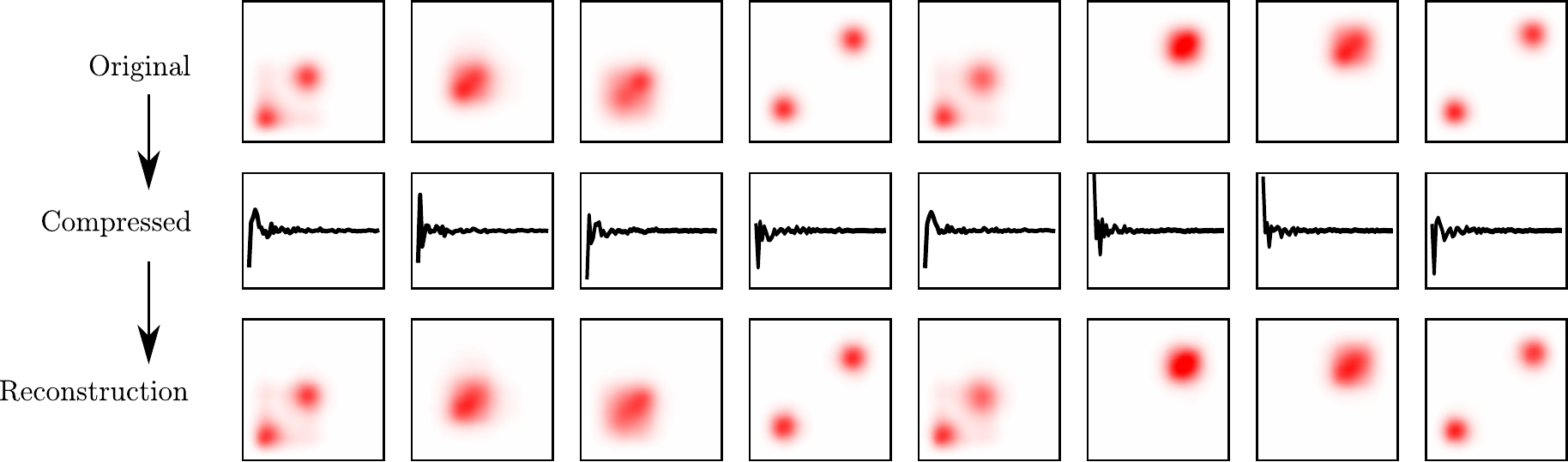}
\caption{The CAE ($N^2 \rightarrow N \rightarrow N^2$) being applied to eight example 1-RDMs (illustrated in figure \ref{fig:cae}). The top row shows the original exact 1-RDM in each case (with the same axis as figure \ref{fig:data}). The second row shows the encoded representation of $N$ values in each case. The final row shows the decoder being applied to the compressed data, reconstructing the original $N^2$ 1-RDM to a mean average error of $1.7 \times 10^{-3}$ (a.u.) over the data-set.}
\label{fig:p_dae}
\end{figure*}

% Applying a CAE to the 62 point density matrix, learns symmetry, can learn non-linearity. Can compress from N2 to N.
The PCA is unable to perform the reduction of elements $N^2 \rightarrow N$ because it imposes linearity. Without this constraint, we know this mapping is in principle possible as the $N^2$ elements of the 1-RDM is defined by only $N$, for example from the external potential or charge density. We now transcend this request for linearity by applying a CAE to the 1-RDM for the 62-point data-set, where we set the number of values in the bottleneck layer to be $512$. Applying PCA to the bottleneck data we further reduce to $N=62$. This yields the final mapping of the model to be $N^2 \rightarrow N \rightarrow N^2$ as desired. We find that the model can reconstruct the input to a mean average error of $1.7 \times 10^{-3}$ a.u. (average error of each $\gamma_{ij}$) Figure \ref{fig:p_dae} illustrates the CAE being applied to eight example systems.
Now we have various machine learning models encoding both linear and non-linear constraints for the 1-RDM, and we can utilise what has been learned to construct approximations to the functional $\gamma[n]$.

\section{Learning Functionals}
\subsection{Feature Engineering} \label{sec:feateng}
% What is feature engineering
Before moving on to deep learning the 1-RDM functional, we first investigate to what extent we can assist machine learning models with pre-existing knowledge of the density matrix in the simplest possible case. The cornerstone of this process is \emph{feature engineering}. Any appropriately complex neural network can brute-force correct predictions, but in order to obtain an efficient model it is necessary to determine the best way in which the data should be presented to the machine.

% Feature engineering from the Hubbard model
Let us start by considering the functional $\gamma[n]$ for a two points system containing two electrons of opposite spin. As the 1-RDM has the charge density along its diagonal, and is symmetric, this problem reduces to finding the function 
\begin{equation}
\gamma_{21}(\gamma_{11},\gamma_{22}).
\end{equation}
The universal approximation theorem \cite{hornik1991approximation} guarantees that a MLP with a sufficiently large hidden layer can fit any function. In order to ensure only a small hidden layer is needed, we perform feature engineering. 
Let us start from the two-points Hamiltonian diagonalized by the \texttt{iDEA} code (see section \ref{sec:dataset}):
\begin{align*}
\hat{H} &= -t\sum_{\sigma\in\{\uparrow,\downarrow\}}(\crG{1}\deG{2}+\crG{2}\deG{1})+U\sum_{i=1}^2\nuU{i}\nuD{i}+\\
&+U'(\nuU{1}\nuD{2}+\nuD{1}\nuU{2})+\sum_{i=1}^2{v_i\nuG{i}}
\end{align*}
The term $U'$ corresponds to the repulsion of the electrons when populating different sites. The distance in between the points has been appropriately tuned in order to make this term negligible with respect to the on-site repulsion, so that the system can be modelled as an inhomogeneous Hubbard-dimer model:
\begin{equation}
\hat{H} = -t\sum_{\sigma\in\{\uparrow,\downarrow\}}(\crG{1}\deG{2}+\crG{2}\deG{1})+U\sum_{i=1}^2\nuU{i}\nuD{i}+\sum_{i=1}^2{v_i\nuG{i}}
\end{equation}
Starting from this Hamiltonian, let us define an adimensional quantity named interaction strength
$$u=\frac{U}{4t}.$$
This number represents the relative importance of the on-site repulsion with respect to the kinetic term.  
By performing a variational constrained minimization of the Hamiltonian \cite{saubanere2016interaction,tows2011lattice} it is possible to extract the desired functional in the two limiting cases of non-interacting electrons :
\begin{equation}\label{eq:gamma21_non}
\gamma_{21}^0=\gamma_{21}(u=0)=\sqrt{\gamma_{11}(2-\gamma_{11})},
\end{equation}
and of strongly-interacting electrons:
\begin{equation}\label{eq:gamma21_inf}
\gamma^{\infty}_{21}=\gamma_{21} (u\rightarrow\infty)=
\begin{cases}
\sqrt{2(\gamma_{11}-1)(2-\gamma_{11})}  & \text{if $\gamma_{11}\geq 1$}\\
\sqrt{2\gamma_{11}(1-\gamma_{11})} & \text{if $\gamma_{11}< 1$.} \\ 
\end{cases}
\end{equation}
In terms of the variable $\gamma_m=\mathrm{min}\{\gamma_{11},2-\gamma_{11}\}$ this becomes
\begin{equation}
\gamma^{\infty}_{2,1}=\sqrt{2\gamma_{m}|1-\gamma_m|}.
\end{equation}
This allows us to drastically reduce the complexity of the network needed for fitting the data. The relations can be written as 
\begin{equation}
\gamma_{21}(x_1,x_2)=x_1^{\omega_1}x_2^{\omega_2}=e^{\omega_1\log{x_1}+\omega_2\log{x_2}+b},
\end{equation}
where $\omega_i=1/2$ and $b=0$
\begin{equation*}
(x_1,x_2) =
\begin{cases}
(2\gamma_m,1-\gamma_m)  & \text{if strongly-interacting}\\
(\gamma_{11},\gamma_{22}) & \text{if non-interacting}.
\end{cases}
\end{equation*}
We can then define
\begin{equation}
O_{P}=f^{\sigma}(\sum_k \omega_k\hat{x}_k+b),
\end{equation} 
where $f^{\sigma}(x)=e^x$, $\sum_k \omega_k\hat{x}_k$ is the weighted sum of the inputs, that are defined to be $\hat{x}=\log{x}$ and the bias is given by $b$. $O_{P}$, in the presented form, is the output of a perceptron, which is the simplest neural network, as being composed by one single neuron. This is termed a logarithmic perceptron \cite{hines1996logarithmic}. 
It is important to note that a brute force MLP could always yield an equally accurate result, but it would require a large hidden layer of many perceptrons. In contrast this model needs only one. The computational burden has been reduced to a three parameter model to be fitted by the logarithm of the original input data.

% The logarithmic perceptron
We train the logarithmic perceptron using the mean square error loss function and the Adam optimizer with a learning rate of $10^{-3}$. The bias has been initialized to 0 and a norm-2 bias regularizer with a coefficient of 10 has been introduced in order to highly penalize any value of the bias different from zero. The average parameters of 20 training sessions, computed both for the interacting and for the non interacting case are reported in table \ref{tab:log_perc}. As expected, the machine has learned that the bias is negligible\footnote{$b=o(\omega)$ since it is smaller than the precision with which the value of $\omega$ is known.} with respect to the $\omega$-parameters, that have been estimated to be $\omega\simeq0.5$. While for the non-interacting case the result is exact, being the non-interacting condition exactly reproducible, the strongly interacting case only approximately matches the infinitely interacting case, being this condition a limit.
\begin{table}[]
\centering
\begin{tabular}{l||lll}
& $\omega_1$ &  $\omega_2$ & $b$   \\
$\gamma^0$ & $0.5000\pm 0.0002$ & $0.5001\pm0.0001$ &  $(2.0\pm0.1)10^{-6}$ \\
$\gamma^{\infty}$ & $0.480\pm 0.003$& $0.480\pm0.002$ &  $(-4\pm 8)10^{-5}$ \\
\end{tabular}
\caption{Result of the fitting procedure using the logarithmic perceptron as an average over 20 example training sessions, along with the corresponding uncertainty.}
\label{tab:log_perc}
\end{table}

% summary
We have determined that two analytical limits can be encoded in an engineered minimally complex architecture. In this small system, this yields a network that is vastly simpler than a brute force MLP. The logarithmic perceptron is engineered to optimally describe the relationship in between the variables and so is a candidate building block to construct neural network models for finding the desired functionals when more than two grid-points are concerned. This is because it could be possible to take advantage of the capability of this perceptron to introduce the correct non-linearity, while possibly allowing to physically interpret the final architecture as a nested combination of Hubbard dimers for modelling systems with a higher number of grid-points \cite{hines1996logarithmic}.

\subsection{A perturbative approach}
% What is this approach
In section \ref{sec:feateng} we have shown that the two point system can be modeled as a Hubbard dimer, and we have given the explicit functional form of the off-diagonal term in the two limiting cases. This then defines a domain between these two cases.
% COMMENT 1: v$-representability domain \cite{saubanere2016interaction}\luc{do we need the concept of "representability" in the following? see also comment later on the "unphysical". The two limiting cases refer to the interaction, right? usually we talk about representability wrt the density, the potential.....is here really the interaction meant? If yes, maybe this solves your problem of "unphysical" below. If not, is it needed? If we do not know, I would say as little as possible at this stage, for lack of time to be rigorous. }. In this section we will extract an analytical form of the $\gamma_{21}$-functional in the whole range $u\in(0,\infty)$. The model proposed will be characterized by a dominant term, deduced analytically, while a feed-forward neural network will be used for completing the theory by fitting the correction.

% Defining the approach
The starting point is to express the equations \ref{eq:gamma21_non} and \ref{eq:gamma21_inf} in terms of the variable $\gamma_m$:
\begin{align}
\gamma_{21}^0(\gamma_m) &= \sqrt{2\gamma_m-1\gamma_{m}^2}\\
\gamma_{21}^{\infty}(\gamma_m) &= \sqrt{2\gamma_m-2\gamma_{m}^2}
\end{align}
By observing the structure of these laws, we postulate that the functional form at intermediate values of the interaction strength can be written as
\begin{equation}
\gamma_{2,1}(\gamma_m,u) = \sqrt{2\gamma_m-\chi(\gamma_m,u)\gamma_m^2}.
\end{equation}
The point $\gamma_m=1$ is a special value for the Hubbard dimer model since it corresponds to the point in which the value of the density at the two sites is the same. This can only occur when the dimer is symmetric ($v_1=v_2$). 
Performing the diagonalization of the Hamiltonian of the symmetric dimer, the value of the off-diagonal term of the 1-RDM as a function of the interaction strength is found to be 
\begin{equation}
\gamma_{21}(\gamma_m=1,u) = -\frac{u-\sqrt{u^2+1}}{1+u(u-\sqrt{u^2+1})}.
\end{equation}
This relation fixes the value of the  $\chi$ function at the symmetric point: 
\begin{equation}
\chi(\gamma_m=1,u)= 2-[\gamma_{2,1}(\gamma_m=1,u)]^2.
\end{equation}
Apart from this, nothing obvious can be said about the $\gamma_m$ dependence of the $\chi$ function. However, we choose to write it as 
\begin{equation}
\begin{split}
\chi(\gamma_m,u) &=\chi(\gamma_m=1,u)+\Delta\chi(\gamma_m,u) \\
&=\chi^{(0)}(u)+\Delta\chi(\gamma_m,u)
\label{eq:chi0}
\end{split}
\end{equation}
and the following functional constraints must necessarily be true
\begin{equation}
\begin{cases}
\Delta\chi(\gamma_{m}=1,u) = 0\\
\Delta\chi(\gamma_{m},u=0) = 0\\
\Delta\chi(\gamma_{m},u\rightarrow\infty) = 0.
\end{cases}
\end{equation}
The first constraint, by definition, is valid whatever $u$ in the symmetry point $(\gamma_{m}=1)$.
%It holds as a direct consequence of the previous definition of $\chi^{(0)}(u)$. 
The second and the third constraints are due to the fact that $\chi^{(0)}(u=0)=1$ and $\lim_{u\rightarrow\infty}\chi^{(0)}(u)=2$. Since this must be true for all the values of $\gamma_m$, the correction must be zero.

% Testing the chi zero model
Considering that the correction vanishes at both the borders of our domain, and also at the symmetry point, and that a crossings of two any curves of the off-diagonal term for different values of the interaction strength should not occur, one would expect the correction to be a perturbation of the $\chi^{(0)}$-model. 
% COMMENT 2: \ac{unphysical:Notation:\begin{itemize}
% \item $|a_1|^2$: proba double occupancy first site
% \item $|a_3|^2$: proba double occupancy second site
% \item $|a_2|^2$: proba one per each
% $$\end{itemize}
% $|\Psi\rangle = a_1|1\uparrow1\downarrow\rangle+a_2\frac{1}{\sqrt{2}}(|1\uparrow2\downarrow\rangle+|1\downarrow2\uparrow\rangle+a_3|2\uparrow2\downarrow\rangle)$. Let us assume by absurd that there exists two different values of the interaction strength $u_h$ and $u_l$, $u_h>u_l$ for which at the same density profile, uniquely defined by $\gamma_{1,1}$, the $\gamma_{2,1}$-terms are the same. this means that $\gamma_{2,1}^h = \sqrt{2}(a_1^ha_2^h+a_2^ha_3^h)$ is equal to $\gamma_{2,1}^l = \sqrt{2}(a_1^la_2^l+a_2^la_3^l)$ and that $2|a_1^h|^2+|a_2^h|^2=2|a_1^l|^2+|a_2^l|^2$ and $2|a_3^h|^2+|a_2^h|^2=2|a_3^l|^2+|a_2^l|^2$. This implies that all of the terms must be equal, i.e. that for two different values of the interaction strength we can have the same ground state. So this also means, for instance, that the probability for two electrons to be in the same site ($|a_1|^2$ or $|a_3|^2$) is the same for different values of the interaction strength, which is non-sense. This does not apply to the boundaries. a density equal to zero corresponds to a value of the potential not accessible. In that case there's no difference in between the interaction strengths since the electrons are forced to live in the only existing site.}

\begin{figure}[htbp]
\centering
\includegraphics[width=1.0\linewidth]{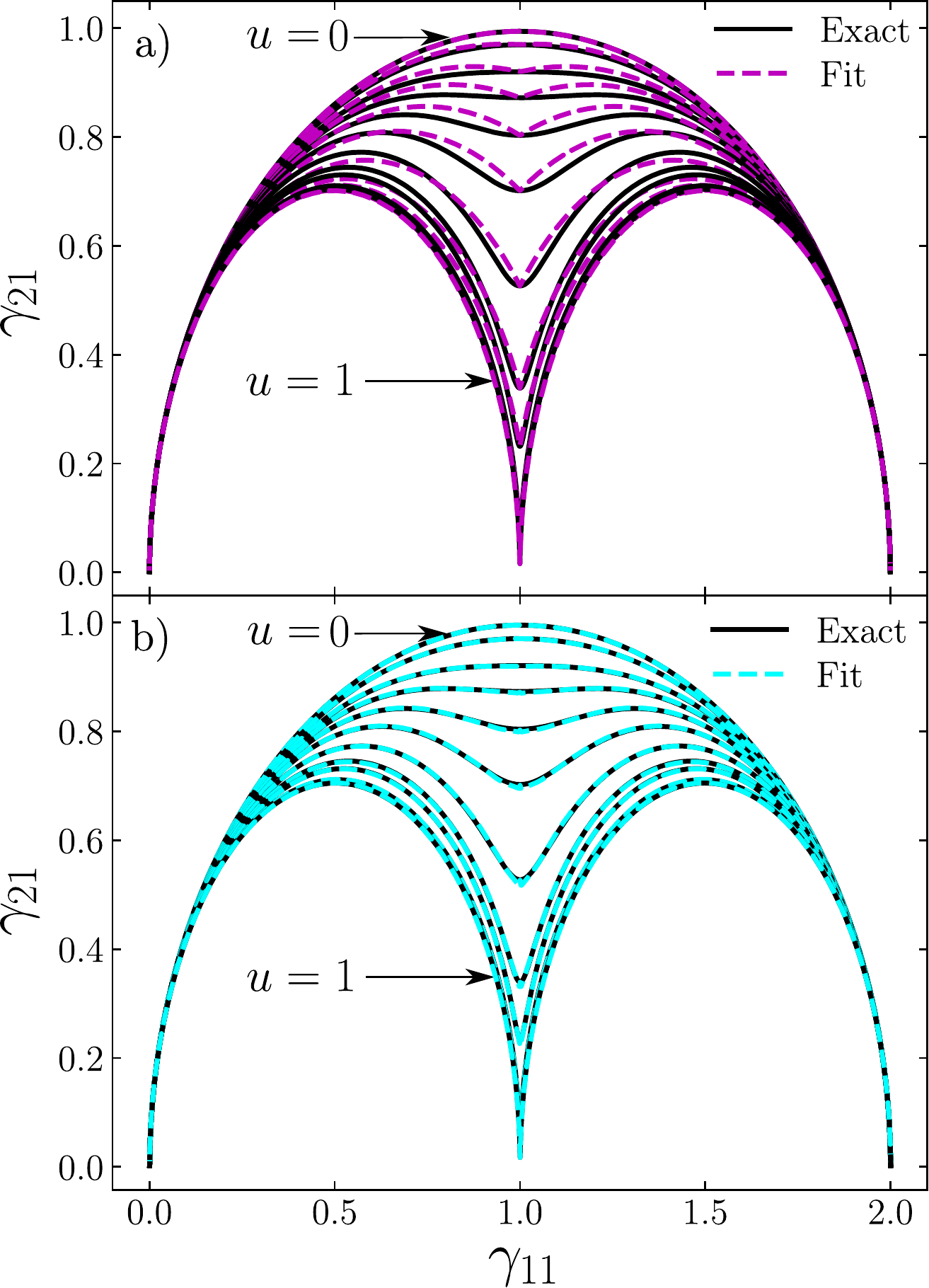}
\caption{Panel (a) shows the performance of the zeroth order $\chi$-model for a range of interaction strengths $u$ from 0 to 1, in comparison to the exact case. Panel (b) shows the $\chi$-model with the machine learned correction. This demonstrates that the machine has learned a significant improvement.}
\label{fig:chi}
\end{figure}

In figure \ref{fig:chi}(a) we directly compare the $\chi^{(0)}$-model with the exact .

% Learning the correction
As this has verified that the correction is indeed a perturbation of the proposed model, we now employ a neural network architecture to determine this correction. We generate an additional data-set containing 600,000 couples $(\hat{\gamma}_m,\hat{u})$. For a range of values of $u$ different potential landscapes have been defined and the density matrix has been computed. The corresponding values of $\Delta\hat{\chi}(\gamma_m,u)\gamma_m^2$ have been used as labels to be learned in the regression procedure. The quantities have been redefined as $\hat{x} = 10x$ and the factor 10 has been introduced to ensure the data is of favorable scale for working in non-linearity with the selected activation functions. This activation function is chosen to be the hyperbolic tangent as it is capable of reaching negative values. A reasonable choice for the number of neurons in the three layers has been found to be $(12,12,16)$. Due to the simplicity of the model some details of the correction are missed in the fitting, in particular the vanishing of the correction at the symmetry point and the vanishing of the correction in the non-interacting limit. Rather than increasing the complexity of the network we have preferred to impose this functional requirement by multiplying the prediction by two exponential corrections. The equation of the correction reads
\begin{equation}
    \Delta\chi(\gamma_m,u)\gamma_m^2\simeq\frac{O(\hat{\gamma}_m,\hat{u})}{10}(1-e^{-\frac{1-\gamma_m}{\lambda_g}})(1-e^{-\frac{u}{\lambda_u}})
\end{equation}
where $O(\hat{\gamma}_m,\hat{u})$ is the prediction of the network while $\lambda_g = 0.001$ and $\lambda_u = 0.004$ are two numerical coefficients appropriately chosen. Figure \ref{fig:chi}(b) compares the inclusion of this correction to the exact result, showing a significant increase in accuracy. 

% Transition
Up to this moment we have used machine learning tools in order to enhance our theoretical models and to learn them. We will now move to larger grid-points systems, using the capability of the machine to learn from the data for the construction of approximate functionals.

\subsection{Learning functionals from constraints} \label{sec:PCAfunc}
% Functionals can be split into linear and non-linear
We will now move to applying the insights into constraints we gained in sections \ref{sec:PCAcharge} and \ref{sec:pcadm} to develop the functional $\gamma[n]$ for the 62 point data-set, and we will then benchmark the resulting estimations of the 1-RDMs  against the exact ones. The functional we desire can be split into a linear (L), and non-linear (NL) term:
\begin{equation}\label{eq:lin_nonlin}
\gamma[n] = \gamma_\mathrm{L}[n] + \gamma_\mathrm{NL}[n].
\end{equation}
In the following we will explicitly perform this linear decomposition. While the linear term will be presented as an explicit functional of the density, the non-linear one will be treated as a perturbation, and will be deep learned in section \ref{sec:DAE}.

% PCA can be used to construct linear part domain specific
In section \ref{sec:pcadm} we found that with PCA, due to additional structural constraints, the principal components contained non-zero values in both the diagonal and non-diagonal elements. In particular, the diagonal of the 1-RDM components had a significant correspondence to the density components (see figure \ref{fig:p_pca2}). For a given number of grid points, this correspondence holds for the first $\nu$ components, where this value is determined by analysing the PCA components. The purpose of this section is to exploit this correspondence in order to find a linear approximation of ${\gamma}[n]$. We will begin by formalizing the connection in between the two data-sets introduced in sections \ref{sec:PCAcharge} and \ref{sec:pcadm}. We will first introduce the principal component decomposition of the 1-RDM in the $\mathcal{B}_e$-basis introduced in section \ref{sec:pca} and then, starting from the matrix representation of the density data-set, we will express in formulas the content of figure \ref{fig:p_pca2}. The starting point is the expression of the 1-RDM components in terms of the known projections of the principal components onto the basis defining the entries of the matrix (see section \ref{sec:pca}).
\begin{align*}
\gamma_{i,j} &= \LA e_{r[i,j]}|\gamma\RA =\\
&= \LA e_{r[i,j]}|\gamma_0\RA+\LA e_{r[i,j]}|\Tilde{\gamma}\RA= \\
&= (\gamma_0)_{i,j}+\sum_{k,r'=1}^{N^2}\LA p_k|e_{r'}\RA\LA e_{r'}|\Tilde{\gamma}\RA\LA e_{r[i,j]}|p_k\RA
\end{align*}
Where the $\LA e_r|p_i\RA$ coefficients are known after the evaluation of the eigenvectors of the covariance matrix. \newline
Before proceeding, we introduce the density vectors:
\begin{equation}
\underline{n} = [n(x_1),0,\cdots,0,n(x_2),0,\cdots,n(x_N)]\in\mathbb{R}^{N^2}
\end{equation}
and the corresponding data-set
\begin{equation}
\underline{\underline{P}}=\begin{bmatrix}
\underline{n}^{(1)}\\
\vdots\\
\underline{n}^{(T)}
\end{bmatrix}.
\end{equation}
Performing a PCA on this data-set allows one to determine the average density $|n_0\RA $, the decomposition $|n\RA = |\Tilde{n}\RA+|n_0\RA$ and the corresponding principal components $$\{|p^{n}_i\RA\}_{i=1}^{N^2}.$$ There are $N^2$ components since they must form an orthonormal basis of the vector-space. However, only the first $N-1$ principal components will be informative due to the sparseness of the object defined and due to the normalization of the density (see section \ref{sec:pcadm}).
%COMMENT 3: \luc{why $N^2$ and not $N$?}
%\ac{because the principal components need to form an orthonormal basis of the space in which the vectors in the data-set live. We can discuss about it but I am quite sure about the fact that, if there is no variation along some direction (like in this case, in which there are always zero components along $N^2-N$ directions), then, as you say, the first $N$ principal components will be informative because the associates eigenvalues of the correlation matrix will be non-zero. For what concerns the remaining principal components/directions they must exist because we need a complete basis but the only criterion they need to satisfy is to be orthogonal to the previous ones. What counts is that their associated singular value is negligible, which means they are not informative.}
By direct comparison of the principal directions in the two data-sets (see figure \ref{fig:p_pca1}), it is possible to observe that the first $\nu$ principal directions derived from the $\Gamma$ data-set can be put approximately in a scaled one-to-one correspondence with the first $\nu$ principal directions of the sub-data-set $P$. 
In particular, let us define the modified principal directions and let us normalize them $$|q_i\rangle = \sum_{j=1}^{N}\langle e_{r[j,j]}|p_i\rangle|e_{r[j,j]}\RA\hspace{0.3cm}i=1,\cdots,N-1 \rightarrow |{q}^{n}_i\RA = \frac{1}{\sqrt{\LA q_i|{q}_i\RA}}|q_i\RA.$$
This defines an orthonormal basis for describing the diagonals of the matrices in the ensemble. When the matrices under analysis are such that the non-vanishing off-diagonal terms are mainly the ones closer to the corresponding non-vanishing diagonal terms, it must be true that $|q^n_i\rangle \simeq \pm |p^{n}_i\RA$.\footnote{The $\pm$ is due to the possible differing convention of the arbitrary directions in the PCA.} Where the equality has been observed to be exact for the first $\nu$ principal components since the presence in the density matrix of the off-diagonal terms leads to a reduction in priority of the variation along the density. In fact, the main variations in the density are also those more strikingly characterizing the 1-RDM. However, from the $(\nu+1)-th$ component on, while the PCA on the density can provide more details on the remaining changes in the density, orthogonal to the previous ones, the PCA on the density matrix, starts describing the variations along the off-diagonal terms and the mapping in between the two breaks down since the details on the density, being less evident than the ones of the off-diagonal terms, are contained in a diluted way in the remaining components.

% derivation of PCA functional
We will now move to determine the actual expression of the linear functional. First, we separate the mean-adjusted 1-RDM in four terms, distinguishing the diagonal from the off-diagonal contributions and taking into account the different information content of the first $\nu$ principal component with respect to the remaining ones. We will then define a basis orthonormal in the subspace of the densities while carrying off-diagonal information. This definition will allow to approximate the dominant contribution to the exact functional $\gamma[n]$. Let us start by writing $\Tilde{\gamma}$ as the sum of four contributions:
\begin{align*}
|\Tilde{\gamma}\RA &=|\Tilde{n}_{\leq \nu}\RA +|\Tilde{n}_{> \nu}\RA+ |\delta{\gamma}_{\leq \nu}\RA+|\delta{\gamma}_{>\nu}\RA
\end{align*}
These four terms correspond to the shifted density reconstructed with the first $\nu$ principal components, to the shifted density reconstructed with the remaining principal components, to the off-diagonal terms obtainable with the first $\nu$ principal components, from now on termed \emph{free off-diagonal terms}, and to the remaining off-diagonal contributions. Considering or not these terms corresponds to different levels of approximation.
% COMMENT 4: \ac{I am referring to the 20 points data-set. The 90\% thing is a rule of thumb that works in my case but I don't know if it works in the 64 data points. I used this because it was easier for me to explain the whole thing but please, change this with your SVD thing. The only important thing is to describe how we choose the $\nu$ number and the fact that from that value on the connection of the principal directions of the density and of the 1-RDM falls apart but still it is not a big deal for the reason you pointed out}
For our specific data-set it has been shown that the moment in which the one-to-one mapping stops to hold corresponds to the number of principal components $\nu$ at which the cumulative sum of the explained variance ratio reaches a value of $0.91$. For this reason, neglecting the term $|\Tilde{\gamma}_{>\nu}\RA=|\Tilde{n}_{> \nu}\RA+|\delta\gamma_{> \nu}\RA$ 
will be considered as a reasonable first order approximation, and we will be able to focus on the remaining two terms.
This having been said, let us define a new set of vectors:
$$|q_i^{\gamma}\RA\doteq\frac{1}{\sqrt{\LA q_i|q_i\RA}}|p_i\RA \hspace{0.3cm}i=1,\cdots,N-1 \hspace{1cm} \LA q^n_i|q_j^{\gamma}\RA = \delta_{i,j}$$
These vectors contain the $|q_i^n\RA$ ones in them and follow their normalization. Thanks to their orthogonality, if the one-to-one mapping were valid for all the first $N-1$ components, they would be a complete basis in the densities-subspace while varying the information on the free off-diagonal terms in the components not shared with the basis $\{|q_i^{n}\RA\}_{i=1}^{N-1}$. Even if the mapping is valid only for the first $\nu$ principal components, this basis allows nonetheless for the construction of the approximate 1-RDM carrying some information on the off-diagonal behavior while its diagonal corresponds to the density reconstruction obtained by looking at its $\nu$ most remarkable features. Let us add and subtract this term in $\gamma_{i,j}$:
\begin{align*}
\gamma_{i,j} &=  (\gamma_0)_{i,j} + \LA e_{r[i,j]}|\Tilde{n}_{\leq\nu}\RA + \LA e_{r[i,j]}|\delta{\gamma}_{\leq\nu}\RA+ \LA e_{r[i,j]}|\Tilde{\gamma}_{>\nu}\RA+\\
&+\sum_{k=1}^{\nu}\LA e_{r[i,j]}|q_k^{\gamma}\RA\LA q_k^{\gamma} |\Tilde{n}\RA-\sum_{k=1}^{\nu}\LA e_{r[i,j]}|q_k^{\gamma}\RA\LA q_k^{\gamma} |\Tilde{n}\RA=\\
&= (\gamma_0)_{i,j} + \sum_{k=1}^{\nu}\LA e_{r[i,j]}|q_k^{\gamma}\RA\LA q_k^{\gamma} |\Tilde{n}\RA +(\delta\gamma)_{i,j}
\end{align*}
where 
\begin{equation*}
(\delta\gamma)_{i,j} = \LA e_{r[i,j]}|\Tilde{n}_{\leq\nu}\RA+(\delta{\gamma}_{\leq\nu})_{i,j}-\sum_{k=1}^{\nu}\LA e_{r[i,j]}|q_k^{\gamma}\RA\LA q_k^{\gamma} |\Tilde{n}\RA +(\Tilde{\gamma}_{> \nu})_{i,j}
\end{equation*}
This last quantity must be itself a functional of the density, where the functional relation is non-linear and unknown. This having been done, the functional is now expressed in the form presented in equation \ref{eq:lin_nonlin}.
%\begin{equation}
%    \gamma[n] = \gamma_\mathrm{L}[n] + \gamma_\mathrm{NL}[n].
%\end{equation}
% COMMENT 5 \luc{Overall, this derivation looks correct, but it is not easy to follow and it seems to me that some things are redundant. In particular, eq. 33 and eq. 30 are the same, so it would be enough to refer backwards. Now, what do we really need for the derivation and analysis? If you define the $\tilde \gamma$ vector as the sum of 4 terms, as you do above and which one can always do, and if you define the $q^1$ basis and the projection of the $\tilde n<$ term in this basis, and if you then add and subtract this projection to your sum of 4 terms, you directly have the final result, right? Of course you want to discuss it, but are you sure that you need for this everything you have introduced, e.g. the $q^0$ etc? Could you check whether you need all definitions, and all lines in the derivation?}
Considering the non-linear part, the analysis on the principal values legitimates us to neglect the term $(\Tilde{\gamma}_{> \nu})_{i,j}$. For what concerns the remaining contribution it is expected to be small since the basis $\{|q_k^{\gamma}\RA\}_{k=1}^{N-1}$ has been defined with the exact intent of privileging the exact restoration of the density, being the biggest contribution, while estimating the free off-diagonal terms.
By neglecting the $\gamma_\mathrm{NL}[n]=\delta\gamma$ term and by writing the resulting expression in terms of the know projections of the principal components onto the $\mathcal{B}_e$-basis, the linear functional is obtained
\begin{align*}
\gamma_{i,j}&= (\gamma_0)_{i,j} + \sum_{r': \LA e_{r'}|\Tilde{n}\RA\neq 0}\sum_{k=1}^{\nu}\LA e_{r[i,j]}|q_k^{\gamma}\RA\LA q_k^{\gamma} |e_{r'}\RA\LA e_{r'}|\Tilde{n}\RA\\
&= (\gamma_0)_{i,j} + \sum_{s=1}^N(n(x_s)-n_0(x_s))\sum_{k=1}^{\nu}\LA e_{r[i,j]}|q_k^{\gamma}\RA\LA q_k^{\gamma} |e_{r'[s,s]}\RA
\end{align*}
\begin{equation}
\gamma_{i,j}[n]= (\gamma_0)_{i,j}+\sum_{s=1}^{N} (n(x_s)-n_0(x_s))\sum_{k=1}^{\nu}\LA q_k^1|e_{r'[s,s]}\RA\LA e_{r[i,j]}|q_k^1\RA
\end{equation}
Writing this in the position basis for our 1-dimensional data-set, we arrive at our approximate functional, we term this \emph{the PCA functional}:
\begin{equation}
\gamma_\mathrm{L}^\mathrm{PCA}[n(\tilde{x})](x,x') = \gamma_{0}(x,x') + \hat{P}\hat{P}^{-1}_{\nu}(n(\tilde{x}) - n_{0}(\tilde{x}))(x,x'),
\end{equation}
where $\gamma_{0}$ is the average density matrix from the data-set (due to the PCA convention to transform between mean-adjusted data), $\hat{P}$ is the PCA 1-RDM transformation, and $(n - n_0)$ is the mean adjusted charge-density. $\hat{P}^{-1}_{\nu}$ is the diagonal-only inverse PCA transform. This takes a mean-adjusted density, and returns the $\nu$ PCA components of $\gamma$ that when transformed to real space will contain that density along its diagonal. This chooses our PCA components of the density matrix so they must have our given density along its diagonal. When the PCA transform is applied to give the density matrix in real space we also obtain off diagonal elements linearly approximated by this inverse transformation. This inverse has some very small eigenvalues due to some of the off-diagonal principal components of $\gamma$ containing small diagonal values. We use singular value decomposition to remove these eigenvalue in order to perform the inverse.

% Comparison to Taylor expansion
This approximate functional can be thought of as a \emph{domain specific} linear expansion, akin to a Taylor expansion. It is domain specific in two ways; first that the PCA orders components by variance, where the neglected terms are the smallest possible by definition, and so is engineered for an optimal linear approximation. Secondly, that the region of which the expansion is accurate has been specified by a data-set of systems of interest. This can be made analogous to domain specificity in image compression: An autoencoder can be trained to yield optimal compression on a specific data-set (domain) of images (for example faces). If this were instead trained on all possible images, one would recover something akin to JPEG compression, and hence autoencoders are thought as domain specific image compression. In this way of thinking, this approximate functional is a \emph{domain specific} linear expansion, as it has been trained on a representative data-set of systems, which is a small subset of all possible systems.

\begin{figure}[htbp]
\centering
\includegraphics[width=1.0\linewidth]{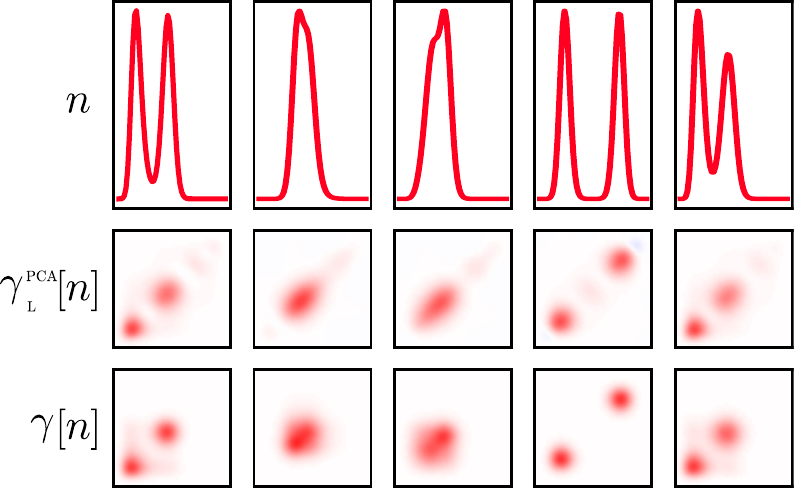}
\caption{Evaluating our PCA linear functional for nine sample systems from the 62-point data-set. The axis are the same as in figure \ref{fig:data}. The first row shows each system's charge densities. The second row shows the PCA functional being applied to each of the charge densities. The third row shows the exact density matrix corresponding to each of the densities. By taking the average mean percentage difference taken over the entire data-set we find that the linear functional takes account of 64.41\% of the whole density matrix, leaving only the remaining to be deep learned.}
\label{fig:func_constraint}
\end{figure}

% Performance on the data-set
Figure \ref{fig:func_constraint} shows the application the the PCA functional to nine example systems from the 62-point data-set, with $\nu=8$. We find that a significant contribution (on average 64.41\%) of the off-diagonal elements can be described by this linear functional. This leaves only the non-linear term to be learned. In section \ref{sec:DAE} we will go beyond this linear term using a deep learning model.

\subsection{Denoising Autoencoders} \label{sec:DAE}
% What are DAE?
We will now move to approximating $\gamma[n]$ for the 62-point systems using denoising autoencoders (DAEs). DAEs are convolutional autoencoders used to perform noise reduction in image processing\cite{Goodfellow-et-al-2016}. Usually CAEs are trained to reconstruct their input exactly, but if noise is applied to the data-set, it can instead be trained to construct the clean data from the data with noise added. When given a novel noisy image it can reconstruct the image with the noise removed. We propose that DAEs can be used to develop functionals if we treat the difference between an approximate 1-RDM and the exact to be noise.
\FloatBarrier
\begin{figure}[htbp]
\centering
\includegraphics[width=1.0\linewidth]{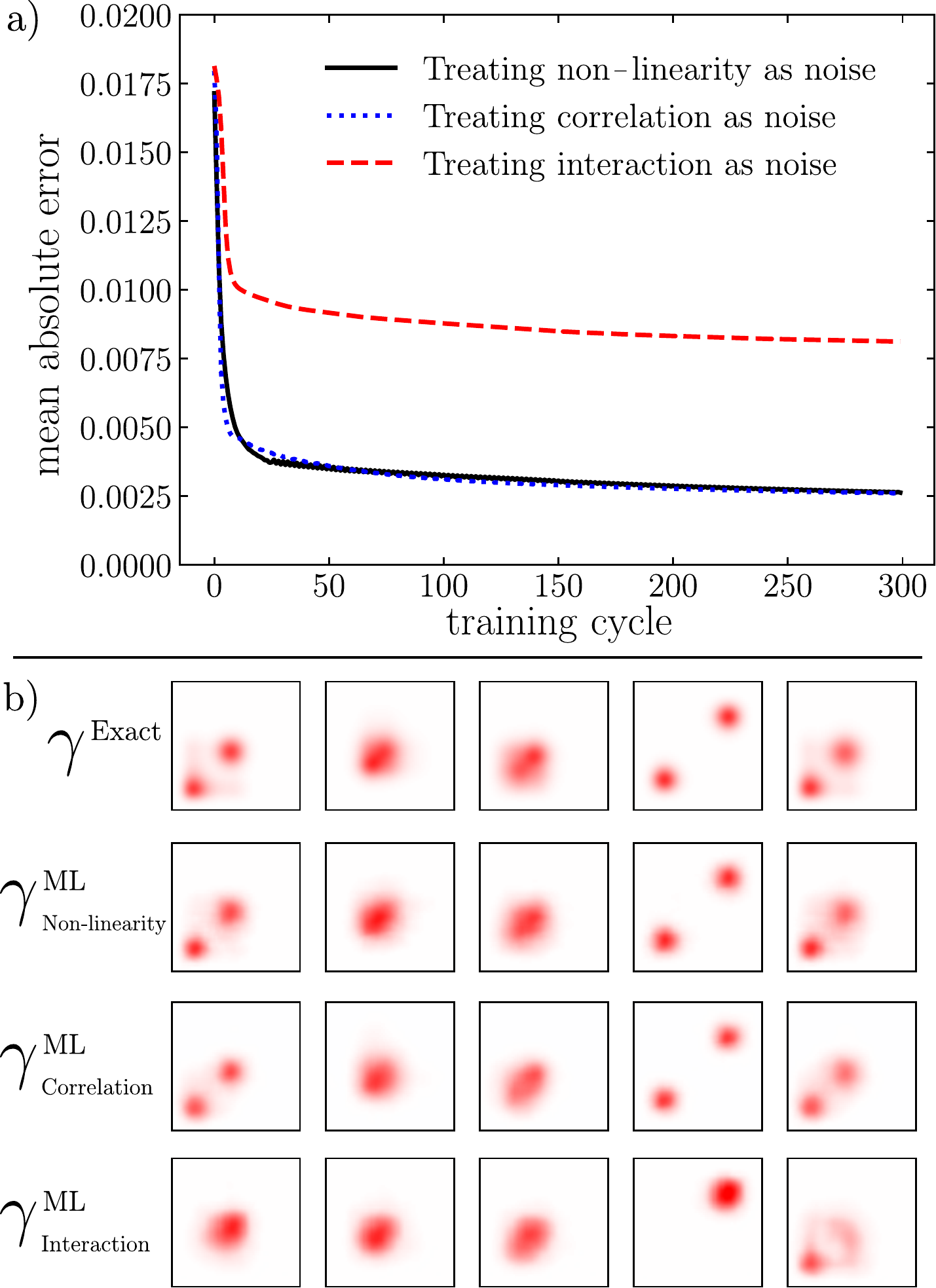}
\caption{Contrasting the different quantities we can treat as noise when training a DAE. Panel (a) shows the mean absolute error as a function of training cycle (where the DAE has seen every training sample once). Treating the neglect of non-linearity and correlation as noise converges to a mean absolute error $2.6 \times 10^{-3}$ (a.u), whereas treating the neglect of the entirety of interaction as noise converges to $8.2 \times 10^{-3}$ (a.u).  Panel (b) shows the application of these three deep learning methods to five example test systems in comparison to the exact.}
\label{fig:denoise}
\end{figure}
\FloatBarrier

% We can begin with different choices of what to ML (Linearity, interaction, correlation)
 We have several candidates of what we can consider noise:
\begin{itemize}
\item By training our DAE to reconstruct the exact density matrix from the PCA functional introduced in section \ref{sec:PCAfunc}, we are considering the neglect of non-linearity as noise.
\item By training our DAE to reconstruct the exact density matrix from the purely non-interacting, we are considering the neglect of the whole Coulomb interaction as noise. 
\item By training our DAE to reconstruct the exact density matrix from the UHF, we are considering as noise the neglect of correlation beyond that simulated by the symmetry breaking.
\end{itemize}

% Comparison of training in each case
By training a DAE in each of these three cases we can see which phenomena are most amenable to be deep learned in this fashion. In figure \ref{fig:denoise} (a) we show, for equivalent DAE architectures, the mean absolute error in the predictions as a function of the training cycle. This shows that the neglect of interaction is the least applicable to be treated as noise. Treating the neglect of non-linearity and neglect of correlation effects as noise are largely equally as applicable, yielding a final mean absolute error of $2.6 \times 10^{-3}$ (a.u.)  (in comparison to the UHF approximation to the 1-RDM itself has a mean absolute error of $2.0 \times 10^{-2}$ (a.u.) over the data-set).  In figure \ref{fig:denoise} (b) we illustrate the prediction of each of the approximate deep learning methods for five example systems in comparison to the exact. It is clear to see that machine learning the interaction itself is considerable less amenable to machine learning than either the correlation effects beyond UHF and non-linearity. And therefore, combining the linear functional obtained in section \ref{sec:PCAfunc} with a denoising encoder, yields a accurate approximation to the functional $\gamma[n]$.

\section{Conclusions}
In conclusion, we have shown that insights into the one-body reduced density matrix (1-RDM) can be gained using a variety of machine learning methods. We show that by employing a large data-set of 1-RDMs, the machine can learn the constraints underlying the data. Linear constraints are determined by principal component analysis (PCA). The PCA illustrates that using a constrained class of external potentials, in addition to the usual smoothness constraints, leads to additional linear constraints in the charge density. Subsequently, non-linear constraints can be learned from convolutional autoencoders (CAEs). We show that these constraints can be utilised to build approximations to the 1-RDM as a functional of the charge density. The PCA can be used to construct the linear part of the functional utilizing linear constraints. Subsequently, the neglect of the non-linear contribution can be considered as noise, which in turn can be rectified using a denoising autoencoder (DAE). This approach yields accurate density matrices as functions of the charge density when applied to exactly solvable model systems. We compare what quantity can best be treated as noise in this way, when building functionals using DAEs, and hence which unknown term is most amenable to machine learning. We find that the treatment of interaction is considerably more difficult than non-linearity or correlation effects beyond unrestricted Hartree-Fock (UHF). We also show how existing knowledge of the density matrix can be used to guide machine learning techniques, in particular the construction of networks using logarithmic neurons, which is a candidate to assemble more complex machine learning strategies. This two-way transfer of knowledge between existing approaches and machine learning strategies is expected to help both the analytic design of new functionals, and numerical approaches to materials computation based on machine learning.

\section*{Conflicts of interest}
%In accordance with our policy on \href{http://www.rsc.org/journals-books-databases/journal-authors-reviewers/author-responsibilities/#code-of-conduct}{Conflicts of interest} please ensure that a conflicts of interest statement is included in your manuscript here.  Please note that this statement is required for all submitted manuscripts.  If no conflicts exist, please state that ``
There are no conflicts to declare.

\section*{Acknowledgements}
We thank the University of York for computational resources.

%%%END OF MAIN TEXT%%%
\balance
%%%REFERENCES%%%
\bibliography{preprint} %You need to replace "rsc" on this line with the name of your .bib file

\begin{thebibliography}{58}%
\makeatletter
\providecommand \@ifxundefined [1]{%
 \@ifx{#1\undefined}
}%
\providecommand \@ifnum [1]{%
 \ifnum #1\expandafter \@firstoftwo
 \else \expandafter \@secondoftwo
 \fi
}%
\providecommand \@ifx [1]{%
 \ifx #1\expandafter \@firstoftwo
 \else \expandafter \@secondoftwo
 \fi
}%
\providecommand \natexlab [1]{#1}%
\providecommand \enquote  [1]{``#1''}%
\providecommand \bibnamefont  [1]{#1}%
\providecommand \bibfnamefont [1]{#1}%
\providecommand \citenamefont [1]{#1}%
\providecommand \href@noop [0]{\@secondoftwo}%
\providecommand \href [0]{\begingroup \@sanitize@url \@href}%
\providecommand \@href[1]{\@@startlink{#1}\@@href}%
\providecommand \@@href[1]{\endgroup#1\@@endlink}%
\providecommand \@sanitize@url [0]{\catcode `\\12\catcode `\$12\catcode
  `\&12\catcode `\#12\catcode `\^12\catcode `\_12\catcode `\%12\relax}%
\providecommand \@@startlink[1]{}%
\providecommand \@@endlink[0]{}%
\providecommand \url  [0]{\begingroup\@sanitize@url \@url }%
\providecommand \@url [1]{\endgroup\@href {#1}{\urlprefix }}%
\providecommand \urlprefix  [0]{URL }%
\providecommand \Eprint [0]{\href }%
\providecommand \doibase [0]{http://dx.doi.org/}%
\providecommand \selectlanguage [0]{\@gobble}%
\providecommand \bibinfo  [0]{\@secondoftwo}%
\providecommand \bibfield  [0]{\@secondoftwo}%
\providecommand \translation [1]{[#1]}%
\providecommand \BibitemOpen [0]{}%
\providecommand \bibitemStop [0]{}%
\providecommand \bibitemNoStop [0]{.\EOS\space}%
\providecommand \EOS [0]{\spacefactor3000\relax}%
\providecommand \BibitemShut  [1]{\csname bibitem#1\endcsname}%
\let\auto@bib@innerbib\@empty
%</preamble>
\bibitem [{\citenamefont {Hohenberg}\ and\ \citenamefont
  {Kohn}(1964)}]{Hohenberg-Kohn}%
  \BibitemOpen
  \bibfield  {author} {\bibinfo {author} {\bibfnamefont {P.}~\bibnamefont
  {Hohenberg}}\ and\ \bibinfo {author} {\bibfnamefont {W.}~\bibnamefont
  {Kohn}},\ }\href@noop {} {\bibfield  {journal} {\bibinfo  {journal} {Phys.
  Rev.}\ }\textbf {\bibinfo {volume} {136}},\ \bibinfo {pages} {864B} (\bibinfo
  {year} {1964})}\BibitemShut {NoStop}%
\bibitem [{\citenamefont {Coleman}(1963)}]{Coleman1963}%
  \BibitemOpen
  \bibfield  {author} {\bibinfo {author} {\bibfnamefont {A.~J.}\ \bibnamefont
  {Coleman}},\ }\href {\doibase 10.1103/RevModPhys.35.668} {\bibfield
  {journal} {\bibinfo  {journal} {Rev. Mod. Phys.}\ }\textbf {\bibinfo {volume}
  {35}},\ \bibinfo {pages} {668} (\bibinfo {year} {1963})}\BibitemShut
  {NoStop}%
\bibitem [{\citenamefont {Gilbert}(1975)}]{Gilbert1975}%
  \BibitemOpen
  \bibfield  {author} {\bibinfo {author} {\bibfnamefont {T.~L.}\ \bibnamefont
  {Gilbert}},\ }\href {\doibase 10.1103/PhysRevB.12.2111} {\bibfield  {journal}
  {\bibinfo  {journal} {Phys. Rev. B}\ }\textbf {\bibinfo {volume} {12}},\
  \bibinfo {pages} {2111} (\bibinfo {year} {1975})}\BibitemShut {NoStop}%
\bibitem [{\citenamefont {Levy}(1979)}]{Levy1979}%
  \BibitemOpen
  \bibfield  {author} {\bibinfo {author} {\bibfnamefont {M.}~\bibnamefont
  {Levy}},\ }\href {\doibase 10.1073/pnas.76.12.6062} {\bibfield  {journal}
  {\bibinfo  {journal} {Proceedings of the National Academy of Sciences}\
  }\textbf {\bibinfo {volume} {76}},\ \bibinfo {pages} {6062} (\bibinfo {year}
  {1979})},\ \Eprint
  {http://arxiv.org/abs/https://www.pnas.org/content/76/12/6062.full.pdf}
  {https://www.pnas.org/content/76/12/6062.full.pdf} \BibitemShut {NoStop}%
\bibitem [{\citenamefont {Valone}(1980)}]{Valone1980}%
  \BibitemOpen
  \bibfield  {author} {\bibinfo {author} {\bibfnamefont {S.~M.}\ \bibnamefont
  {Valone}},\ }\href {\doibase 10.1063/1.440249} {\bibfield  {journal}
  {\bibinfo  {journal} {The Journal of Chemical Physics}\ }\textbf {\bibinfo
  {volume} {73}},\ \bibinfo {pages} {1344} (\bibinfo {year} {1980})},\ \Eprint
  {http://arxiv.org/abs/https://doi.org/10.1063/1.440249}
  {https://doi.org/10.1063/1.440249} \BibitemShut {NoStop}%
\bibitem [{\citenamefont {Pernal}\ and\ \citenamefont
  {Giesbertz}(2016)}]{Pernal2016}%
  \BibitemOpen
  \bibfield  {author} {\bibinfo {author} {\bibfnamefont {K.}~\bibnamefont
  {Pernal}}\ and\ \bibinfo {author} {\bibfnamefont {K.~J.~H.}\ \bibnamefont
  {Giesbertz}},\ }\enquote {\bibinfo {title} {Reduced density matrix functional
  theory (rdmft) and linear response time-dependent rdmft (td-rdmft)},}\ in\
  \href {\doibase 10.1007/128_2015_624} {\emph {\bibinfo {booktitle}
  {Density-Functional Methods for Excited States}}},\ \bibinfo {editor} {edited
  by\ \bibinfo {editor} {\bibfnamefont {N.}~\bibnamefont {Ferr{\'e}}}, \bibinfo
  {editor} {\bibfnamefont {M.}~\bibnamefont {Filatov}}, \ and\ \bibinfo
  {editor} {\bibfnamefont {M.}~\bibnamefont {Huix-Rotllant}}}\ (\bibinfo
  {publisher} {Springer International Publishing},\ \bibinfo {address} {Cham},\
  \bibinfo {year} {2016})\ pp.\ \bibinfo {pages} {125--183}\BibitemShut
  {NoStop}%
\bibitem [{\citenamefont {Lathiotakis}\ \emph {et~al.}(2007)\citenamefont
  {Lathiotakis}, \citenamefont {Helbig},\ and\ \citenamefont
  {Gross}}]{Lathiotakis2007}%
  \BibitemOpen
  \bibfield  {author} {\bibinfo {author} {\bibfnamefont {N.~N.}\ \bibnamefont
  {Lathiotakis}}, \bibinfo {author} {\bibfnamefont {N.}~\bibnamefont {Helbig}},
  \ and\ \bibinfo {author} {\bibfnamefont {E.~K.~U.}\ \bibnamefont {Gross}},\
  }\href {\doibase 10.1103/PhysRevB.75.195120} {\bibfield  {journal} {\bibinfo
  {journal} {Phys. Rev. B}\ }\textbf {\bibinfo {volume} {75}},\ \bibinfo
  {pages} {195120} (\bibinfo {year} {2007})}\BibitemShut {NoStop}%
\bibitem [{\citenamefont {Piris}(2017)}]{Piris2017}%
  \BibitemOpen
  \bibfield  {author} {\bibinfo {author} {\bibfnamefont {M.}~\bibnamefont
  {Piris}},\ }\href {\doibase 10.1103/PhysRevLett.119.063002} {\bibfield
  {journal} {\bibinfo  {journal} {Phys. Rev. Lett.}\ }\textbf {\bibinfo
  {volume} {119}},\ \bibinfo {pages} {063002} (\bibinfo {year}
  {2017})}\BibitemShut {NoStop}%
\bibitem [{\citenamefont {Schilling}(2018)}]{Schilling2018}%
  \BibitemOpen
  \bibfield  {author} {\bibinfo {author} {\bibfnamefont {C.}~\bibnamefont
  {Schilling}},\ }\href {\doibase 10.1063/1.5080088} {\bibfield  {journal}
  {\bibinfo  {journal} {The Journal of Chemical Physics}\ }\textbf {\bibinfo
  {volume} {149}},\ \bibinfo {pages} {231102} (\bibinfo {year} {2018})},\
  \Eprint {http://arxiv.org/abs/https://doi.org/10.1063/1.5080088}
  {https://doi.org/10.1063/1.5080088} \BibitemShut {NoStop}%
\bibitem [{\citenamefont {Giesbertz}\ \emph {et~al.}(2018)\citenamefont
  {Giesbertz}, \citenamefont {Uimonen},\ and\ \citenamefont {van
  Leeuwen}}]{Giesbertz2018}%
  \BibitemOpen
  \bibfield  {author} {\bibinfo {author} {\bibfnamefont {K.~J.~H.}\
  \bibnamefont {Giesbertz}}, \bibinfo {author} {\bibfnamefont {A.-M.}\
  \bibnamefont {Uimonen}}, \ and\ \bibinfo {author} {\bibfnamefont
  {R.}~\bibnamefont {van Leeuwen}},\ }\href {\doibase
  10.1140/epjb/e2018-90279-1} {\bibfield  {journal} {\bibinfo  {journal} {The
  European Physical Journal B}\ }\textbf {\bibinfo {volume} {91}},\ \bibinfo
  {pages} {282} (\bibinfo {year} {2018})}\BibitemShut {NoStop}%
\bibitem [{\citenamefont {Gritsenko}\ \emph {et~al.}(2005)\citenamefont
  {Gritsenko}, \citenamefont {Pernal},\ and\ \citenamefont
  {Baerends}}]{doi:10.1063/1.1906203}%
  \BibitemOpen
  \bibfield  {author} {\bibinfo {author} {\bibfnamefont {O.}~\bibnamefont
  {Gritsenko}}, \bibinfo {author} {\bibfnamefont {K.}~\bibnamefont {Pernal}}, \
  and\ \bibinfo {author} {\bibfnamefont {E.~J.}\ \bibnamefont {Baerends}},\
  }\href {\doibase 10.1063/1.1906203} {\bibfield  {journal} {\bibinfo
  {journal} {The Journal of Chemical Physics}\ }\textbf {\bibinfo {volume}
  {122}},\ \bibinfo {pages} {204102} (\bibinfo {year} {2005})},\ \Eprint
  {http://arxiv.org/abs/https://doi.org/10.1063/1.1906203}
  {https://doi.org/10.1063/1.1906203} \BibitemShut {NoStop}%
\bibitem [{\citenamefont {Sharma}\ \emph {et~al.}(2008)\citenamefont {Sharma},
  \citenamefont {Dewhurst}, \citenamefont {Lathiotakis},\ and\ \citenamefont
  {Gross}}]{PhysRevB.78.201103}%
  \BibitemOpen
  \bibfield  {author} {\bibinfo {author} {\bibfnamefont {S.}~\bibnamefont
  {Sharma}}, \bibinfo {author} {\bibfnamefont {J.~K.}\ \bibnamefont
  {Dewhurst}}, \bibinfo {author} {\bibfnamefont {N.~N.}\ \bibnamefont
  {Lathiotakis}}, \ and\ \bibinfo {author} {\bibfnamefont {E.~K.~U.}\
  \bibnamefont {Gross}},\ }\href {\doibase 10.1103/PhysRevB.78.201103}
  {\bibfield  {journal} {\bibinfo  {journal} {Phys. Rev. B}\ }\textbf {\bibinfo
  {volume} {78}},\ \bibinfo {pages} {201103} (\bibinfo {year}
  {2008})}\BibitemShut {NoStop}%
\bibitem [{\citenamefont {Kohn}\ and\ \citenamefont {Sham}(1965)}]{Kohn-Sham}%
  \BibitemOpen
  \bibfield  {author} {\bibinfo {author} {\bibfnamefont {W.}~\bibnamefont
  {Kohn}}\ and\ \bibinfo {author} {\bibfnamefont {L.}~\bibnamefont {Sham}},\
  }\href@noop {} {\bibfield  {journal} {\bibinfo  {journal} {Phys. Rev.}\
  }\textbf {\bibinfo {volume} {140}},\ \bibinfo {pages} {1133A} (\bibinfo
  {year} {1965})}\BibitemShut {NoStop}%
\bibitem [{\citenamefont {Goedecker}\ and\ \citenamefont
  {Umrigar}(1998)}]{PhysRevLett.81.866}%
  \BibitemOpen
  \bibfield  {author} {\bibinfo {author} {\bibfnamefont {S.}~\bibnamefont
  {Goedecker}}\ and\ \bibinfo {author} {\bibfnamefont {C.~J.}\ \bibnamefont
  {Umrigar}},\ }\href {\doibase 10.1103/PhysRevLett.81.866} {\bibfield
  {journal} {\bibinfo  {journal} {Phys. Rev. Lett.}\ }\textbf {\bibinfo
  {volume} {81}},\ \bibinfo {pages} {866} (\bibinfo {year} {1998})}\BibitemShut
  {NoStop}%
\bibitem [{\citenamefont {Hollingsworth}\ \emph {et~al.}(2018)\citenamefont
  {Hollingsworth}, \citenamefont {Li}, \citenamefont {Baker},\ and\
  \citenamefont {Burke}}]{doi:10.1063/1.5025668}%
  \BibitemOpen
  \bibfield  {author} {\bibinfo {author} {\bibfnamefont {J.}~\bibnamefont
  {Hollingsworth}}, \bibinfo {author} {\bibfnamefont {L.}~\bibnamefont {Li}},
  \bibinfo {author} {\bibfnamefont {T.~E.}\ \bibnamefont {Baker}}, \ and\
  \bibinfo {author} {\bibfnamefont {K.}~\bibnamefont {Burke}},\ }\href
  {\doibase 10.1063/1.5025668} {\bibfield  {journal} {\bibinfo  {journal} {The
  Journal of Chemical Physics}\ }\textbf {\bibinfo {volume} {148}},\ \bibinfo
  {pages} {241743} (\bibinfo {year} {2018})},\ \Eprint
  {http://arxiv.org/abs/https://doi.org/10.1063/1.5025668}
  {https://doi.org/10.1063/1.5025668} \BibitemShut {NoStop}%
\bibitem [{\citenamefont {Carleo}\ \emph {et~al.}(2019)\citenamefont {Carleo},
  \citenamefont {Cirac}, \citenamefont {Cranmer}, \citenamefont {Daudet},
  \citenamefont {Schuld}, \citenamefont {Tishby}, \citenamefont
  {Vogt-Maranto},\ and\ \citenamefont {Zdeborov\'a}}]{RevModPhys.91.045002}%
  \BibitemOpen
  \bibfield  {author} {\bibinfo {author} {\bibfnamefont {G.}~\bibnamefont
  {Carleo}}, \bibinfo {author} {\bibfnamefont {I.}~\bibnamefont {Cirac}},
  \bibinfo {author} {\bibfnamefont {K.}~\bibnamefont {Cranmer}}, \bibinfo
  {author} {\bibfnamefont {L.}~\bibnamefont {Daudet}}, \bibinfo {author}
  {\bibfnamefont {M.}~\bibnamefont {Schuld}}, \bibinfo {author} {\bibfnamefont
  {N.}~\bibnamefont {Tishby}}, \bibinfo {author} {\bibfnamefont
  {L.}~\bibnamefont {Vogt-Maranto}}, \ and\ \bibinfo {author} {\bibfnamefont
  {L.}~\bibnamefont {Zdeborov\'a}},\ }\href {\doibase
  10.1103/RevModPhys.91.045002} {\bibfield  {journal} {\bibinfo  {journal}
  {Rev. Mod. Phys.}\ }\textbf {\bibinfo {volume} {91}},\ \bibinfo {pages}
  {045002} (\bibinfo {year} {2019})}\BibitemShut {NoStop}%
\bibitem [{\citenamefont {Snyder}\ \emph {et~al.}(2012)\citenamefont {Snyder},
  \citenamefont {Rupp}, \citenamefont {Hansen}, \citenamefont {M\"uller},\ and\
  \citenamefont {Burke}}]{PhysRevLett.108.253002}%
  \BibitemOpen
  \bibfield  {author} {\bibinfo {author} {\bibfnamefont {J.~C.}\ \bibnamefont
  {Snyder}}, \bibinfo {author} {\bibfnamefont {M.}~\bibnamefont {Rupp}},
  \bibinfo {author} {\bibfnamefont {K.}~\bibnamefont {Hansen}}, \bibinfo
  {author} {\bibfnamefont {K.-R.}\ \bibnamefont {M\"uller}}, \ and\ \bibinfo
  {author} {\bibfnamefont {K.}~\bibnamefont {Burke}},\ }\href {\doibase
  10.1103/PhysRevLett.108.253002} {\bibfield  {journal} {\bibinfo  {journal}
  {Phys. Rev. Lett.}\ }\textbf {\bibinfo {volume} {108}},\ \bibinfo {pages}
  {253002} (\bibinfo {year} {2012})}\BibitemShut {NoStop}%
\bibitem [{\citenamefont {Li}\ \emph {et~al.}(2016{\natexlab{a}})\citenamefont
  {Li}, \citenamefont {Baker}, \citenamefont {White},\ and\ \citenamefont
  {Burke}}]{PhysRevB.94.245129}%
  \BibitemOpen
  \bibfield  {author} {\bibinfo {author} {\bibfnamefont {L.}~\bibnamefont
  {Li}}, \bibinfo {author} {\bibfnamefont {T.~E.}\ \bibnamefont {Baker}},
  \bibinfo {author} {\bibfnamefont {S.~R.}\ \bibnamefont {White}}, \ and\
  \bibinfo {author} {\bibfnamefont {K.}~\bibnamefont {Burke}},\ }\href
  {\doibase 10.1103/PhysRevB.94.245129} {\bibfield  {journal} {\bibinfo
  {journal} {Phys. Rev. B}\ }\textbf {\bibinfo {volume} {94}},\ \bibinfo
  {pages} {245129} (\bibinfo {year} {2016}{\natexlab{a}})}\BibitemShut
  {NoStop}%
\bibitem [{\citenamefont {Li}\ \emph {et~al.}(2016{\natexlab{b}})\citenamefont
  {Li}, \citenamefont {Snyder}, \citenamefont {Pelaschier}, \citenamefont
  {Huang}, \citenamefont {Niranjan}, \citenamefont {Duncan}, \citenamefont
  {Rupp}, \citenamefont {Müller},\ and\ \citenamefont
  {Burke}}]{doi:10.1002/qua.25040}%
  \BibitemOpen
  \bibfield  {author} {\bibinfo {author} {\bibfnamefont {L.}~\bibnamefont
  {Li}}, \bibinfo {author} {\bibfnamefont {J.~C.}\ \bibnamefont {Snyder}},
  \bibinfo {author} {\bibfnamefont {I.~M.}\ \bibnamefont {Pelaschier}},
  \bibinfo {author} {\bibfnamefont {J.}~\bibnamefont {Huang}}, \bibinfo
  {author} {\bibfnamefont {U.-N.}\ \bibnamefont {Niranjan}}, \bibinfo {author}
  {\bibfnamefont {P.}~\bibnamefont {Duncan}}, \bibinfo {author} {\bibfnamefont
  {M.}~\bibnamefont {Rupp}}, \bibinfo {author} {\bibfnamefont {K.-R.}\
  \bibnamefont {Müller}}, \ and\ \bibinfo {author} {\bibfnamefont
  {K.}~\bibnamefont {Burke}},\ }\href {\doibase 10.1002/qua.25040} {\bibfield
  {journal} {\bibinfo  {journal} {International Journal of Quantum Chemistry}\
  }\textbf {\bibinfo {volume} {116}},\ \bibinfo {pages} {819} (\bibinfo {year}
  {2016}{\natexlab{b}})},\ \Eprint
  {http://arxiv.org/abs/https://onlinelibrary.wiley.com/doi/pdf/10.1002/qua.25040}
  {https://onlinelibrary.wiley.com/doi/pdf/10.1002/qua.25040} \BibitemShut
  {NoStop}%
\bibitem [{\citenamefont {Behler}\ and\ \citenamefont
  {Parrinello}(2007)}]{PhysRevLett.98.146401}%
  \BibitemOpen
  \bibfield  {author} {\bibinfo {author} {\bibfnamefont {J.}~\bibnamefont
  {Behler}}\ and\ \bibinfo {author} {\bibfnamefont {M.}~\bibnamefont
  {Parrinello}},\ }\href {\doibase 10.1103/PhysRevLett.98.146401} {\bibfield
  {journal} {\bibinfo  {journal} {Phys. Rev. Lett.}\ }\textbf {\bibinfo
  {volume} {98}},\ \bibinfo {pages} {146401} (\bibinfo {year}
  {2007})}\BibitemShut {NoStop}%
\bibitem [{\citenamefont {Bart\'ok}\ \emph {et~al.}(2010)\citenamefont
  {Bart\'ok}, \citenamefont {Payne}, \citenamefont {Kondor},\ and\
  \citenamefont {Cs\'anyi}}]{PhysRevLett.104.136403}%
  \BibitemOpen
  \bibfield  {author} {\bibinfo {author} {\bibfnamefont {A.~P.}\ \bibnamefont
  {Bart\'ok}}, \bibinfo {author} {\bibfnamefont {M.~C.}\ \bibnamefont {Payne}},
  \bibinfo {author} {\bibfnamefont {R.}~\bibnamefont {Kondor}}, \ and\ \bibinfo
  {author} {\bibfnamefont {G.}~\bibnamefont {Cs\'anyi}},\ }\href {\doibase
  10.1103/PhysRevLett.104.136403} {\bibfield  {journal} {\bibinfo  {journal}
  {Phys. Rev. Lett.}\ }\textbf {\bibinfo {volume} {104}},\ \bibinfo {pages}
  {136403} (\bibinfo {year} {2010})}\BibitemShut {NoStop}%
\bibitem [{\citenamefont {Moreno}\ \emph {et~al.}(2019)\citenamefont {Moreno},
  \citenamefont {Carleo},\ and\ \citenamefont {Georges}}]{moreno2019deep}%
  \BibitemOpen
  \bibfield  {author} {\bibinfo {author} {\bibfnamefont {J.~R.}\ \bibnamefont
  {Moreno}}, \bibinfo {author} {\bibfnamefont {G.}~\bibnamefont {Carleo}}, \
  and\ \bibinfo {author} {\bibfnamefont {A.}~\bibnamefont {Georges}},\
  }\href@noop {} {\enquote {\bibinfo {title} {Deep learning the hohenberg-kohn
  maps of density functional theory},}\ } (\bibinfo {year} {2019}),\ \Eprint
  {http://arxiv.org/abs/1911.03580} {arXiv:1911.03580 [cond-mat.dis-nn]}
  \BibitemShut {NoStop}%
\bibitem [{\citenamefont {Pozun}\ \emph {et~al.}(2012)\citenamefont {Pozun},
  \citenamefont {Hansen}, \citenamefont {Sheppard}, \citenamefont {Rupp},
  \citenamefont {M\"uller},\ and\ \citenamefont
  {Henkelman}}]{doi:10.1063/1.4707167}%
  \BibitemOpen
  \bibfield  {author} {\bibinfo {author} {\bibfnamefont {Z.~D.}\ \bibnamefont
  {Pozun}}, \bibinfo {author} {\bibfnamefont {K.}~\bibnamefont {Hansen}},
  \bibinfo {author} {\bibfnamefont {D.}~\bibnamefont {Sheppard}}, \bibinfo
  {author} {\bibfnamefont {M.}~\bibnamefont {Rupp}}, \bibinfo {author}
  {\bibfnamefont {K.-R.}\ \bibnamefont {M\"uller}}, \ and\ \bibinfo {author}
  {\bibfnamefont {G.}~\bibnamefont {Henkelman}},\ }\href {\doibase
  10.1063/1.4707167} {\bibfield  {journal} {\bibinfo  {journal} {The Journal of
  Chemical Physics}\ }\textbf {\bibinfo {volume} {136}},\ \bibinfo {pages}
  {174101} (\bibinfo {year} {2012})},\ \Eprint
  {http://arxiv.org/abs/https://doi.org/10.1063/1.4707167}
  {https://doi.org/10.1063/1.4707167} \BibitemShut {NoStop}%
\bibitem [{\citenamefont {McGibbon}\ and\ \citenamefont
  {Pande}(2013)}]{McGibbon2013}%
  \BibitemOpen
  \bibfield  {author} {\bibinfo {author} {\bibfnamefont {R.~T.}\ \bibnamefont
  {McGibbon}}\ and\ \bibinfo {author} {\bibfnamefont {V.~S.}\ \bibnamefont
  {Pande}},\ }\href {\doibase 10.1021/ct400132h} {\bibfield  {journal}
  {\bibinfo  {journal} {Journal of Chemical Theory and Computation}\ }\textbf
  {\bibinfo {volume} {9}},\ \bibinfo {pages} {2900} (\bibinfo {year}
  {2013})}\BibitemShut {NoStop}%
\bibitem [{\citenamefont {McDonagh}\ \emph {et~al.}(2018)\citenamefont
  {McDonagh}, \citenamefont {Silva}, \citenamefont {Vincent},\ and\
  \citenamefont {Popelier}}]{McDonagh2018}%
  \BibitemOpen
  \bibfield  {author} {\bibinfo {author} {\bibfnamefont {J.~L.}\ \bibnamefont
  {McDonagh}}, \bibinfo {author} {\bibfnamefont {A.~F.}\ \bibnamefont {Silva}},
  \bibinfo {author} {\bibfnamefont {M.~A.}\ \bibnamefont {Vincent}}, \ and\
  \bibinfo {author} {\bibfnamefont {P.~L.~A.}\ \bibnamefont {Popelier}},\
  }\href {\doibase 10.1021/acs.jctc.7b01157} {\bibfield  {journal} {\bibinfo
  {journal} {Journal of Chemical Theory and Computation}\ }\textbf {\bibinfo
  {volume} {14}},\ \bibinfo {pages} {216} (\bibinfo {year} {2018})}\BibitemShut
  {NoStop}%
\bibitem [{\citenamefont {Rupp}\ \emph {et~al.}(2012)\citenamefont {Rupp},
  \citenamefont {Tkatchenko}, \citenamefont {M\"uller},\ and\ \citenamefont
  {von Lilienfeld}}]{PhysRevLett.108.058301}%
  \BibitemOpen
  \bibfield  {author} {\bibinfo {author} {\bibfnamefont {M.}~\bibnamefont
  {Rupp}}, \bibinfo {author} {\bibfnamefont {A.}~\bibnamefont {Tkatchenko}},
  \bibinfo {author} {\bibfnamefont {K.-R.}\ \bibnamefont {M\"uller}}, \ and\
  \bibinfo {author} {\bibfnamefont {O.~A.}\ \bibnamefont {von Lilienfeld}},\
  }\href {\doibase 10.1103/PhysRevLett.108.058301} {\bibfield  {journal}
  {\bibinfo  {journal} {Phys. Rev. Lett.}\ }\textbf {\bibinfo {volume} {108}},\
  \bibinfo {pages} {058301} (\bibinfo {year} {2012})}\BibitemShut {NoStop}%
\bibitem [{\citenamefont {Hautier}\ \emph {et~al.}(2010)\citenamefont
  {Hautier}, \citenamefont {Fischer}, \citenamefont {Jain}, \citenamefont
  {Mueller},\ and\ \citenamefont {Ceder}}]{Hautier2010}%
  \BibitemOpen
  \bibfield  {author} {\bibinfo {author} {\bibfnamefont {G.}~\bibnamefont
  {Hautier}}, \bibinfo {author} {\bibfnamefont {C.~C.}\ \bibnamefont
  {Fischer}}, \bibinfo {author} {\bibfnamefont {A.}~\bibnamefont {Jain}},
  \bibinfo {author} {\bibfnamefont {T.}~\bibnamefont {Mueller}}, \ and\
  \bibinfo {author} {\bibfnamefont {G.}~\bibnamefont {Ceder}},\ }\href
  {\doibase 10.1021/cm100795d} {\bibfield  {journal} {\bibinfo  {journal}
  {Chemistry of Materials}\ }\textbf {\bibinfo {volume} {22}},\ \bibinfo
  {pages} {3762} (\bibinfo {year} {2010})}\BibitemShut {NoStop}%
\bibitem [{\citenamefont {Kolb}\ \emph {et~al.}(2017)\citenamefont {Kolb},
  \citenamefont {Lentz},\ and\ \citenamefont {Kolpak}}]{Kolb2017}%
  \BibitemOpen
  \bibfield  {author} {\bibinfo {author} {\bibfnamefont {B.}~\bibnamefont
  {Kolb}}, \bibinfo {author} {\bibfnamefont {L.~C.}\ \bibnamefont {Lentz}}, \
  and\ \bibinfo {author} {\bibfnamefont {A.~M.}\ \bibnamefont {Kolpak}},\
  }\href {\doibase 10.1038/s41598-017-01251-z} {\bibfield  {journal} {\bibinfo
  {journal} {Scientific Reports}\ }\textbf {\bibinfo {volume} {7}},\ \bibinfo
  {pages} {1192} (\bibinfo {year} {2017})}\BibitemShut {NoStop}%
\bibitem [{\citenamefont {Ryczko}\ \emph {et~al.}(2019)\citenamefont {Ryczko},
  \citenamefont {Strubbe},\ and\ \citenamefont
  {Tamblyn}}]{PhysRevA.100.022512}%
  \BibitemOpen
  \bibfield  {author} {\bibinfo {author} {\bibfnamefont {K.}~\bibnamefont
  {Ryczko}}, \bibinfo {author} {\bibfnamefont {D.~A.}\ \bibnamefont {Strubbe}},
  \ and\ \bibinfo {author} {\bibfnamefont {I.}~\bibnamefont {Tamblyn}},\ }\href
  {\doibase 10.1103/PhysRevA.100.022512} {\bibfield  {journal} {\bibinfo
  {journal} {Phys. Rev. A}\ }\textbf {\bibinfo {volume} {100}},\ \bibinfo
  {pages} {022512} (\bibinfo {year} {2019})}\BibitemShut {NoStop}%
\bibitem [{\citenamefont {Sch{\"u}tt}\ \emph {et~al.}(2019)\citenamefont
  {Sch{\"u}tt}, \citenamefont {Gastegger}, \citenamefont {Tkatchenko},
  \citenamefont {M{\"u}ller},\ and\ \citenamefont {Maurer}}]{Schutt2019}%
  \BibitemOpen
  \bibfield  {author} {\bibinfo {author} {\bibfnamefont {K.~T.}\ \bibnamefont
  {Sch{\"u}tt}}, \bibinfo {author} {\bibfnamefont {M.}~\bibnamefont
  {Gastegger}}, \bibinfo {author} {\bibfnamefont {A.}~\bibnamefont
  {Tkatchenko}}, \bibinfo {author} {\bibfnamefont {K.-R.}\ \bibnamefont
  {M{\"u}ller}}, \ and\ \bibinfo {author} {\bibfnamefont {R.~J.}\ \bibnamefont
  {Maurer}},\ }\href {\doibase 10.1038/s41467-019-12875-2} {\bibfield
  {journal} {\bibinfo  {journal} {Nature Communications}\ }\textbf {\bibinfo
  {volume} {10}},\ \bibinfo {pages} {5024} (\bibinfo {year}
  {2019})}\BibitemShut {NoStop}%
\bibitem [{\citenamefont {Schmidt}\ \emph {et~al.}(2019)\citenamefont
  {Schmidt}, \citenamefont {Marques}, \citenamefont {Botti},\ and\
  \citenamefont {Marques}}]{Schmidt2019}%
  \BibitemOpen
  \bibfield  {author} {\bibinfo {author} {\bibfnamefont {J.}~\bibnamefont
  {Schmidt}}, \bibinfo {author} {\bibfnamefont {M.~R.~G.}\ \bibnamefont
  {Marques}}, \bibinfo {author} {\bibfnamefont {S.}~\bibnamefont {Botti}}, \
  and\ \bibinfo {author} {\bibfnamefont {M.~A.~L.}\ \bibnamefont {Marques}},\
  }\href {\doibase 10.1038/s41524-019-0221-0} {\bibfield  {journal} {\bibinfo
  {journal} {npj Computational Materials}\ }\textbf {\bibinfo {volume} {5}},\
  \bibinfo {pages} {83} (\bibinfo {year} {2019})}\BibitemShut {NoStop}%
\bibitem [{\citenamefont {Mills}\ \emph {et~al.}(2017)\citenamefont {Mills},
  \citenamefont {Spanner},\ and\ \citenamefont {Tamblyn}}]{PhysRevA.96.042113}%
  \BibitemOpen
  \bibfield  {author} {\bibinfo {author} {\bibfnamefont {K.}~\bibnamefont
  {Mills}}, \bibinfo {author} {\bibfnamefont {M.}~\bibnamefont {Spanner}}, \
  and\ \bibinfo {author} {\bibfnamefont {I.}~\bibnamefont {Tamblyn}},\ }\href
  {\doibase 10.1103/PhysRevA.96.042113} {\bibfield  {journal} {\bibinfo
  {journal} {Phys. Rev. A}\ }\textbf {\bibinfo {volume} {96}},\ \bibinfo
  {pages} {042113} (\bibinfo {year} {2017})}\BibitemShut {NoStop}%
\bibitem [{\citenamefont {Suzuki}\ \emph {et~al.}(2020)\citenamefont {Suzuki},
  \citenamefont {Nagai},\ and\ \citenamefont {Haruyama}}]{suzuki2020machine}%
  \BibitemOpen
  \bibfield  {author} {\bibinfo {author} {\bibfnamefont {Y.}~\bibnamefont
  {Suzuki}}, \bibinfo {author} {\bibfnamefont {R.}~\bibnamefont {Nagai}}, \
  and\ \bibinfo {author} {\bibfnamefont {J.}~\bibnamefont {Haruyama}},\
  }\href@noop {} {\  (\bibinfo {year} {2020})},\ \Eprint
  {http://arxiv.org/abs/2002.06542} {arXiv:2002.06542 [physics.comp-ph]}
  \BibitemShut {NoStop}%
\bibitem [{\citenamefont {Nagai}\ \emph {et~al.}(2020)\citenamefont {Nagai},
  \citenamefont {Akashi},\ and\ \citenamefont {Sugino}}]{Nagai2020}%
  \BibitemOpen
  \bibfield  {author} {\bibinfo {author} {\bibfnamefont {R.}~\bibnamefont
  {Nagai}}, \bibinfo {author} {\bibfnamefont {R.}~\bibnamefont {Akashi}}, \
  and\ \bibinfo {author} {\bibfnamefont {O.}~\bibnamefont {Sugino}},\ }\href
  {\doibase 10.1038/s41524-020-0310-0} {\bibfield  {journal} {\bibinfo
  {journal} {npj Computational Materials}\ }\textbf {\bibinfo {volume} {6}},\
  \bibinfo {pages} {43} (\bibinfo {year} {2020})}\BibitemShut {NoStop}%
\bibitem [{\citenamefont {Zhou}\ \emph {et~al.}(2019)\citenamefont {Zhou},
  \citenamefont {Wu}, \citenamefont {Chen},\ and\ \citenamefont
  {Chen}}]{Zhou2019}%
  \BibitemOpen
  \bibfield  {author} {\bibinfo {author} {\bibfnamefont {Y.}~\bibnamefont
  {Zhou}}, \bibinfo {author} {\bibfnamefont {J.}~\bibnamefont {Wu}}, \bibinfo
  {author} {\bibfnamefont {S.}~\bibnamefont {Chen}}, \ and\ \bibinfo {author}
  {\bibfnamefont {G.}~\bibnamefont {Chen}},\ }\href {\doibase
  10.1021/acs.jpclett.9b02838} {\bibfield  {journal} {\bibinfo  {journal} {The
  Journal of Physical Chemistry Letters}\ }\textbf {\bibinfo {volume} {10}},\
  \bibinfo {pages} {7264} (\bibinfo {year} {2019})}\BibitemShut {NoStop}%
\bibitem [{\citenamefont {Mezey}(2017)}]{doi:10.1063/1.5012279}%
  \BibitemOpen
  \bibfield  {author} {\bibinfo {author} {\bibfnamefont {P.~G.}\ \bibnamefont
  {Mezey}},\ }\href {\doibase 10.1063/1.5012279} {\bibfield  {journal}
  {\bibinfo  {journal} {AIP Conference Proceedings}\ }\textbf {\bibinfo
  {volume} {1906}},\ \bibinfo {pages} {020001} (\bibinfo {year} {2017})},\
  \Eprint
  {http://arxiv.org/abs/https://aip.scitation.org/doi/pdf/10.1063/1.5012279}
  {https://aip.scitation.org/doi/pdf/10.1063/1.5012279} \BibitemShut {NoStop}%
\bibitem [{\citenamefont {Brockherde}\ \emph {et~al.}(2017)\citenamefont
  {Brockherde}, \citenamefont {Vogt}, \citenamefont {Li}, \citenamefont
  {Tuckerman}, \citenamefont {Burke},\ and\ \citenamefont
  {M{\"u}ller}}]{Brockherde2017}%
  \BibitemOpen
  \bibfield  {author} {\bibinfo {author} {\bibfnamefont {F.}~\bibnamefont
  {Brockherde}}, \bibinfo {author} {\bibfnamefont {L.}~\bibnamefont {Vogt}},
  \bibinfo {author} {\bibfnamefont {L.}~\bibnamefont {Li}}, \bibinfo {author}
  {\bibfnamefont {M.~E.}\ \bibnamefont {Tuckerman}}, \bibinfo {author}
  {\bibfnamefont {K.}~\bibnamefont {Burke}}, \ and\ \bibinfo {author}
  {\bibfnamefont {K.-R.}\ \bibnamefont {M{\"u}ller}},\ }\href {\doibase
  10.1038/s41467-017-00839-3} {\bibfield  {journal} {\bibinfo  {journal}
  {Nature Communications}\ }\textbf {\bibinfo {volume} {8}},\ \bibinfo {pages}
  {872} (\bibinfo {year} {2017})}\BibitemShut {NoStop}%
\bibitem [{\citenamefont {McCulloch}\ and\ \citenamefont
  {Pitts}(1943)}]{mcculloch1943logical}%
  \BibitemOpen
  \bibfield  {author} {\bibinfo {author} {\bibfnamefont {W.~S.}\ \bibnamefont
  {McCulloch}}\ and\ \bibinfo {author} {\bibfnamefont {W.}~\bibnamefont
  {Pitts}},\ }\href@noop {} {\bibfield  {journal} {\bibinfo  {journal} {The
  bulletin of mathematical biophysics}\ }\textbf {\bibinfo {volume} {5}},\
  \bibinfo {pages} {115} (\bibinfo {year} {1943})}\BibitemShut {NoStop}%
\bibitem [{\citenamefont {Lenail}()}]{NNSVG}%
  \BibitemOpen
  \bibfield  {author} {\bibinfo {author} {\bibfnamefont {A.}~\bibnamefont
  {Lenail}},\ }\href@noop {} {\bibinfo  {journal}
  {http://alexlenail.me/NN-SVG/}\ }\BibitemShut {NoStop}%
\bibitem [{\citenamefont {Liou}\ \emph {et~al.}(2014)\citenamefont {Liou},
  \citenamefont {Cheng}, \citenamefont {Liou},\ and\ \citenamefont
  {Liou}}]{liou2014autoencoder}%
  \BibitemOpen
\bibfield  {journal} {  }\bibfield  {author} {\bibinfo {author} {\bibfnamefont
  {C.-Y.}\ \bibnamefont {Liou}}, \bibinfo {author} {\bibfnamefont {W.-C.}\
  \bibnamefont {Cheng}}, \bibinfo {author} {\bibfnamefont {J.-W.}\ \bibnamefont
  {Liou}}, \ and\ \bibinfo {author} {\bibfnamefont {D.-R.}\ \bibnamefont
  {Liou}},\ }\href@noop {} {\bibfield  {journal} {\bibinfo  {journal}
  {Neurocomputing}\ }\textbf {\bibinfo {volume} {139}},\ \bibinfo {pages} {84}
  (\bibinfo {year} {2014})}\BibitemShut {NoStop}%
\bibitem [{\citenamefont {Trivedi}\ \emph {et~al.}(2018)\citenamefont
  {Trivedi}, \citenamefont {Srivastava}, \citenamefont {Mishra}, \citenamefont
  {Shukla},\ and\ \citenamefont {Tiwari}}]{TRIVEDI2018525}%
  \BibitemOpen
  \bibfield  {author} {\bibinfo {author} {\bibfnamefont {A.}~\bibnamefont
  {Trivedi}}, \bibinfo {author} {\bibfnamefont {S.}~\bibnamefont {Srivastava}},
  \bibinfo {author} {\bibfnamefont {A.}~\bibnamefont {Mishra}}, \bibinfo
  {author} {\bibfnamefont {A.}~\bibnamefont {Shukla}}, \ and\ \bibinfo {author}
  {\bibfnamefont {R.}~\bibnamefont {Tiwari}},\ }\href {\doibase
  https://doi.org/10.1016/j.procs.2017.12.068} {\bibfield  {journal} {\bibinfo
  {journal} {Procedia Computer Science}\ }\textbf {\bibinfo {volume} {125}},\
  \bibinfo {pages} {525 } (\bibinfo {year} {2018})},\ \bibinfo {note} {the 6th
  International Conference on Smart Computing and Communications}\BibitemShut
  {NoStop}%
\bibitem [{\citenamefont {Zhang}\ \emph {et~al.}(1990)\citenamefont {Zhang},
  \citenamefont {Itoh}, \citenamefont {Tanida},\ and\ \citenamefont
  {Ichioka}}]{Zhang:90}%
  \BibitemOpen
  \bibfield  {author} {\bibinfo {author} {\bibfnamefont {W.}~\bibnamefont
  {Zhang}}, \bibinfo {author} {\bibfnamefont {K.}~\bibnamefont {Itoh}},
  \bibinfo {author} {\bibfnamefont {J.}~\bibnamefont {Tanida}}, \ and\ \bibinfo
  {author} {\bibfnamefont {Y.}~\bibnamefont {Ichioka}},\ }\href {\doibase
  10.1364/AO.29.004790} {\bibfield  {journal} {\bibinfo  {journal} {Appl.
  Opt.}\ }\textbf {\bibinfo {volume} {29}},\ \bibinfo {pages} {4790} (\bibinfo
  {year} {1990})}\BibitemShut {NoStop}%
\bibitem [{\citenamefont {Jolliffe}(2011)}]{Jolliffe2011}%
  \BibitemOpen
  \bibfield  {author} {\bibinfo {author} {\bibfnamefont {I.}~\bibnamefont
  {Jolliffe}},\ }\enquote {\bibinfo {title} {Principal component analysis},}\
  in\ \href {\doibase 10.1007/978-3-642-04898-2_455} {\emph {\bibinfo
  {booktitle} {International Encyclopedia of Statistical Science}}},\ \bibinfo
  {editor} {edited by\ \bibinfo {editor} {\bibfnamefont {M.}~\bibnamefont
  {Lovric}}}\ (\bibinfo  {publisher} {Springer Berlin Heidelberg},\ \bibinfo
  {address} {Berlin, Heidelberg},\ \bibinfo {year} {2011})\ pp.\ \bibinfo
  {pages} {1094--1096}\BibitemShut {NoStop}%
\bibitem [{\citenamefont {Pearson}(1901)}]{pearson1901liii}%
  \BibitemOpen
  \bibfield  {author} {\bibinfo {author} {\bibfnamefont {K.}~\bibnamefont
  {Pearson}},\ }\href@noop {} {\bibfield  {journal} {\bibinfo  {journal} {The
  London, Edinburgh, and Dublin Philosophical Magazine and Journal of Science}\
  }\textbf {\bibinfo {volume} {2}},\ \bibinfo {pages} {559} (\bibinfo {year}
  {1901})}\BibitemShut {NoStop}%
\bibitem [{\citenamefont {Hodgson}\ \emph {et~al.}(2013)\citenamefont
  {Hodgson}, \citenamefont {Ramsden}, \citenamefont {Chapman}, \citenamefont
  {Lillystone},\ and\ \citenamefont {Godby}}]{PhysRevB.88.241102}%
  \BibitemOpen
  \bibfield  {author} {\bibinfo {author} {\bibfnamefont {M.~J.~P.}\
  \bibnamefont {Hodgson}}, \bibinfo {author} {\bibfnamefont {J.~D.}\
  \bibnamefont {Ramsden}}, \bibinfo {author} {\bibfnamefont {J.~B.~J.}\
  \bibnamefont {Chapman}}, \bibinfo {author} {\bibfnamefont {P.}~\bibnamefont
  {Lillystone}}, \ and\ \bibinfo {author} {\bibfnamefont {R.~W.}\ \bibnamefont
  {Godby}},\ }\href@noop {} {\bibfield  {journal} {\bibinfo  {journal} {Phys.
  Rev. B}\ }\textbf {\bibinfo {volume} {88}},\ \bibinfo {pages} {241102}
  (\bibinfo {year} {2013})}\BibitemShut {NoStop}%
\bibitem [{\citenamefont {Wetherell}\ \emph {et~al.}(2019)\citenamefont
  {Wetherell}, \citenamefont {Hodgson}, \citenamefont {Talirz},\ and\
  \citenamefont {Godby}}]{PhysRevB.99.045129}%
  \BibitemOpen
  \bibfield  {author} {\bibinfo {author} {\bibfnamefont {J.}~\bibnamefont
  {Wetherell}}, \bibinfo {author} {\bibfnamefont {M.~J.~P.}\ \bibnamefont
  {Hodgson}}, \bibinfo {author} {\bibfnamefont {L.}~\bibnamefont {Talirz}}, \
  and\ \bibinfo {author} {\bibfnamefont {R.~W.}\ \bibnamefont {Godby}},\ }\href
  {\doibase 10.1103/PhysRevB.99.045129} {\bibfield  {journal} {\bibinfo
  {journal} {Phys. Rev. B}\ }\textbf {\bibinfo {volume} {99}},\ \bibinfo
  {pages} {045129} (\bibinfo {year} {2019})}\BibitemShut {NoStop}%
\bibitem [{\citenamefont {Elmaslmane}\ \emph {et~al.}(2018)\citenamefont
  {Elmaslmane}, \citenamefont {Wetherell}, \citenamefont {Hodgson},
  \citenamefont {McKenna},\ and\ \citenamefont
  {Godby}}]{PhysRevMaterials.2.040801}%
  \BibitemOpen
  \bibfield  {author} {\bibinfo {author} {\bibfnamefont {A.~R.}\ \bibnamefont
  {Elmaslmane}}, \bibinfo {author} {\bibfnamefont {J.}~\bibnamefont
  {Wetherell}}, \bibinfo {author} {\bibfnamefont {M.~J.~P.}\ \bibnamefont
  {Hodgson}}, \bibinfo {author} {\bibfnamefont {K.~P.}\ \bibnamefont
  {McKenna}}, \ and\ \bibinfo {author} {\bibfnamefont {R.~W.}\ \bibnamefont
  {Godby}},\ }\href {\doibase 10.1103/PhysRevMaterials.2.040801} {\bibfield
  {journal} {\bibinfo  {journal} {Phys. Rev. Materials}\ }\textbf {\bibinfo
  {volume} {2}},\ \bibinfo {pages} {040801} (\bibinfo {year}
  {2018})}\BibitemShut {NoStop}%
\bibitem [{\citenamefont {Hodgson}\ \emph {et~al.}(2014)\citenamefont
  {Hodgson}, \citenamefont {Ramsden}, \citenamefont {Durrant},\ and\
  \citenamefont {Godby}}]{PhysRevB.90.241107}%
  \BibitemOpen
  \bibfield  {author} {\bibinfo {author} {\bibfnamefont {M.~J.~P.}\
  \bibnamefont {Hodgson}}, \bibinfo {author} {\bibfnamefont {J.~D.}\
  \bibnamefont {Ramsden}}, \bibinfo {author} {\bibfnamefont {T.~R.}\
  \bibnamefont {Durrant}}, \ and\ \bibinfo {author} {\bibfnamefont {R.~W.}\
  \bibnamefont {Godby}},\ }\href@noop {} {\bibfield  {journal} {\bibinfo
  {journal} {Phys. Rev. B}\ }\textbf {\bibinfo {volume} {90}},\ \bibinfo
  {pages} {241107} (\bibinfo {year} {2014})}\BibitemShut {NoStop}%
\bibitem [{\citenamefont {Hodgson}\ and\ \citenamefont
  {Wetherell}(2020)}]{PhysRevA.101.032502}%
  \BibitemOpen
  \bibfield  {author} {\bibinfo {author} {\bibfnamefont {M.~J.~P.}\
  \bibnamefont {Hodgson}}\ and\ \bibinfo {author} {\bibfnamefont
  {J.}~\bibnamefont {Wetherell}},\ }\href {\doibase
  10.1103/PhysRevA.101.032502} {\bibfield  {journal} {\bibinfo  {journal}
  {Phys. Rev. A}\ }\textbf {\bibinfo {volume} {101}},\ \bibinfo {pages}
  {032502} (\bibinfo {year} {2020})}\BibitemShut {NoStop}%
\bibitem [{\citenamefont {Wetherell}\ \emph {et~al.}(2018)\citenamefont
  {Wetherell}, \citenamefont {Hodgson},\ and\ \citenamefont
  {Godby}}]{PhysRevB.97.121102}%
  \BibitemOpen
  \bibfield  {author} {\bibinfo {author} {\bibfnamefont {J.}~\bibnamefont
  {Wetherell}}, \bibinfo {author} {\bibfnamefont {M.~J.~P.}\ \bibnamefont
  {Hodgson}}, \ and\ \bibinfo {author} {\bibfnamefont {R.~W.}\ \bibnamefont
  {Godby}},\ }\href {\doibase 10.1103/PhysRevB.97.121102} {\bibfield  {journal}
  {\bibinfo  {journal} {Phys. Rev. B}\ }\textbf {\bibinfo {volume} {97}},\
  \bibinfo {pages} {121102} (\bibinfo {year} {2018})}\BibitemShut {NoStop}%
\bibitem [{\citenamefont {Hodgson}\ \emph {et~al.}(2016)\citenamefont
  {Hodgson}, \citenamefont {Ramsden},\ and\ \citenamefont
  {Godby}}]{PhysRevB.93.155146}%
  \BibitemOpen
  \bibfield  {author} {\bibinfo {author} {\bibfnamefont {M.~J.~P.}\
  \bibnamefont {Hodgson}}, \bibinfo {author} {\bibfnamefont {J.~D.}\
  \bibnamefont {Ramsden}}, \ and\ \bibinfo {author} {\bibfnamefont {R.~W.}\
  \bibnamefont {Godby}},\ }\href@noop {} {\bibfield  {journal} {\bibinfo
  {journal} {Phys. Rev. B}\ }\textbf {\bibinfo {volume} {93}},\ \bibinfo
  {pages} {155146} (\bibinfo {year} {2016})}\BibitemShut {NoStop}%
\bibitem [{\citenamefont {Hodgson}\ \emph {et~al.}(2017)\citenamefont
  {Hodgson}, \citenamefont {Kraisler}, \citenamefont {Schild},\ and\
  \citenamefont {Gross}}]{hodgson2017interatomic}%
  \BibitemOpen
  \bibfield  {author} {\bibinfo {author} {\bibfnamefont {M.~J.}\ \bibnamefont
  {Hodgson}}, \bibinfo {author} {\bibfnamefont {E.}~\bibnamefont {Kraisler}},
  \bibinfo {author} {\bibfnamefont {A.}~\bibnamefont {Schild}}, \ and\ \bibinfo
  {author} {\bibfnamefont {E.~K.}\ \bibnamefont {Gross}},\ }\href@noop {}
  {\bibfield  {journal} {\bibinfo  {journal} {The journal of physical chemistry
  letters}\ }\textbf {\bibinfo {volume} {8}},\ \bibinfo {pages} {5974}
  (\bibinfo {year} {2017})}\BibitemShut {NoStop}%
\bibitem [{\citenamefont {Skelt}\ \emph {et~al.}(2018)\citenamefont {Skelt},
  \citenamefont {Godby},\ and\ \citenamefont {D'Amico}}]{Skelt2018}%
  \BibitemOpen
  \bibfield  {author} {\bibinfo {author} {\bibfnamefont {A.~H.}\ \bibnamefont
  {Skelt}}, \bibinfo {author} {\bibfnamefont {R.~W.}\ \bibnamefont {Godby}}, \
  and\ \bibinfo {author} {\bibfnamefont {I.}~\bibnamefont {D'Amico}},\ }\href
  {\doibase 10.1007/s13538-018-0589-1} {\bibfield  {journal} {\bibinfo
  {journal} {Brazilian Journal of Physics}\ }\textbf {\bibinfo {volume} {48}},\
  \bibinfo {pages} {467} (\bibinfo {year} {2018})}\BibitemShut {NoStop}%
\bibitem [{\citenamefont {Hornik}(1991)}]{hornik1991approximation}%
  \BibitemOpen
  \bibfield  {author} {\bibinfo {author} {\bibfnamefont {K.}~\bibnamefont
  {Hornik}},\ }\href@noop {} {\bibfield  {journal} {\bibinfo  {journal} {Neural
  networks}\ }\textbf {\bibinfo {volume} {4}},\ \bibinfo {pages} {251}
  (\bibinfo {year} {1991})}\BibitemShut {NoStop}%
\bibitem [{\citenamefont {Sauban{\`e}re}\ \emph {et~al.}(2016)\citenamefont
  {Sauban{\`e}re}, \citenamefont {Lepetit},\ and\ \citenamefont
  {Pastor}}]{saubanere2016interaction}%
  \BibitemOpen
  \bibfield  {author} {\bibinfo {author} {\bibfnamefont {M.}~\bibnamefont
  {Sauban{\`e}re}}, \bibinfo {author} {\bibfnamefont {M.~B.}\ \bibnamefont
  {Lepetit}}, \ and\ \bibinfo {author} {\bibfnamefont {G.}~\bibnamefont
  {Pastor}},\ }\href@noop {} {\bibfield  {journal} {\bibinfo  {journal}
  {Physical Review B}\ }\textbf {\bibinfo {volume} {94}},\ \bibinfo {pages}
  {045102} (\bibinfo {year} {2016})}\BibitemShut {NoStop}%
\bibitem [{\citenamefont {T{\"o}ws}\ and\ \citenamefont
  {Pastor}(2011)}]{tows2011lattice}%
  \BibitemOpen
  \bibfield  {author} {\bibinfo {author} {\bibfnamefont {W.}~\bibnamefont
  {T{\"o}ws}}\ and\ \bibinfo {author} {\bibfnamefont {G.}~\bibnamefont
  {Pastor}},\ }\href@noop {} {\bibfield  {journal} {\bibinfo  {journal}
  {Physical Review B}\ }\textbf {\bibinfo {volume} {83}},\ \bibinfo {pages}
  {235101} (\bibinfo {year} {2011})}\BibitemShut {NoStop}%
\bibitem [{\citenamefont {Hines}(1996)}]{hines1996logarithmic}%
  \BibitemOpen
  \bibfield  {author} {\bibinfo {author} {\bibfnamefont {J.~W.}\ \bibnamefont
  {Hines}},\ }in\ \href@noop {} {\emph {\bibinfo {booktitle} {Proceedings of
  the 1996 American Nuclear Society, International Topical Meeting on Nuclear
  Plant Instrumentation, Control and Human-Machine Interface Technologies}}},\
  Vol.~\bibinfo {volume} {1}\ (\bibinfo {organization} {Citeseer},\ \bibinfo
  {year} {1996})\ pp.\ \bibinfo {pages} {235--241}\BibitemShut {NoStop}%
\bibitem [{\citenamefont {Goodfellow}\ \emph {et~al.}(2016)\citenamefont
  {Goodfellow}, \citenamefont {Bengio},\ and\ \citenamefont
  {Courville}}]{Goodfellow-et-al-2016}%
  \BibitemOpen
  \bibfield  {author} {\bibinfo {author} {\bibfnamefont {I.}~\bibnamefont
  {Goodfellow}}, \bibinfo {author} {\bibfnamefont {Y.}~\bibnamefont {Bengio}},
  \ and\ \bibinfo {author} {\bibfnamefont {A.}~\bibnamefont {Courville}},\
  }\href@noop {} {\emph {\bibinfo {title} {Deep Learning}}}\ (\bibinfo
  {publisher} {MIT Press},\ \bibinfo {year} {2016})\ \bibinfo {note}
  {\url{http://www.deeplearningbook.org}}\BibitemShut {NoStop}%
\end{thebibliography}%
\bibliographystyle{apsrev4-1} %the RSC's .bst file
\end{document}